\documentclass[fleqn,usenatbib]{mnras}
\usepackage{newtxtext,newtxmath}

\usepackage[T1]{fontenc}
\usepackage{ae,aecompl}
\usepackage{aas_macros}

\usepackage{graphicx}	
\usepackage{amsmath}	
\usepackage{amssymb}	
\usepackage{natbib}
\usepackage{hyperref}

\usepackage{ulem}

\newcommand{\md}{\mathrm{d}}	

\title[Dynamical Friction in the Presence of Outflows]{Simulation of a Compact Object with Outflows Moving Through a Gaseous Background}

\author[Li et al.]{
Xinyu Li$^{1,2}$\thanks{E-mail: xli@cita.utornoto.ca},
Philip Chang$^3$, 
Yuri Levin$^{4,5}$, Christopher D. Matzner$^{6}$ 
and \newauthor
 \hspace{0pt} Philip J. Armitage$^{5,7}$
\\
$^{1}$Canadian Institute for Theoretical Astrophysics, 60 St George St, Toronto, ON M5R 2M8\\
$^{2}$Perimeter Institute for Theoretical Physics, 31 Caroline Street North, Waterloo, Ontario, Canada, N2L 2Y5\\
$^{3}$Department of Physics, University of Wisconsin-Milwaukee, 3135 North Maryland Avenue, Milwaukee, WI 53211, USA\\
$^4$Center for Theoretical Physics, Department of Physics, Columbia University, New York, NY 10027, USA \\
$^5$Center for Computational Astrophysics, Flatiron Institute, New York, NY 10010, USA\\
$^6$Department of Astronomy and Astrophysics, University of Toronto, 50 St George Street, Toronto, ON M5S 3H4, Canada\\
$^7$Department of Physics and Astronomy, Stony Brook University. Stony Brook, NY 11790, USA
}

\date{Accepted XXX. Received YYY; in original form ZZZ}

\begin{document}
\label{firstpage}
\pagerange{\pageref{firstpage}--\pageref{lastpage}}
\maketitle

\begin{abstract}
A compact object moving relative to surrounding gas accretes material and perturbs the density of gas in its vicinity. In the classical picture of Bondi-Hoyle-Lyttleton accretion, the perturbation takes the form of an overdense wake behind the object, which exerts a dynamical friction drag. We use hydrodynamic simulations to investigate how the accretion rate and strength of dynamical friction are modified by the presence of outflow from the compact object.   
We show that the destruction of the wake by an outflow reduces dynamical friction, and reverses its sign when the outflow is strong enough, in good quantitative agreement with analytic calculations.  
For a strong isotropic outflow, the outcome on scales that we have simulated is a negative dynamical friction, i.e., net acceleration.
For jet-like outflows driven by reprocessed accretion, both the rate of accretion and the magnitude of dynamical friction drop for more powerful jets. 
The accretion rate is strongly intermittent when the jet points to the same direction as the motion of the compact object. 
The dynamical effects of outflows may be important for the evolution of compact objects during the common envelope phase of binary systems, and for accreting compact objects and massive stars encountering AGN discs.
\end{abstract}

\begin{keywords}
accretion, accretion discs -- stars: winds, outflows -- stars: black holes -- stars: neutron
\end{keywords}



\section{Introduction}
\label{sec:introduction}
When a massive object moves through a cloud of gas, its gravity pulls ambient gas towards it leading to accretion. In the standard picture of Bondi-Hoyle-Lyttleton (BHL) accretion  \citep{1939PCPS...35..405H,1944MNRAS.104..273B,2004NewAR..48..843E}, the flow is axisymmetric and gas whose unperturbed streamlines pass inside the Bondi radius \footnote{There are a variety of defnitions of in the literature, but all lead to accretion rates that are essentially the same for cases in this paper.}
\begin{equation}
    R_{ B} = \frac{2GM}{c_s^2+V_*^2}
\end{equation}
is accreted onto the object.
Here $M$ and $V_*$ are the mass and velocity of the object and $c_s$ is the sound speed of the ambient gas.
The total accretion rate, valid for both subsonic and supersonic cases, is given by \citet{1952MNRAS.112..195B} as
\begin{equation}\label{eq:acc_rate}
    \dot{M}_B=\alpha\pi R_{ B}^2\rho_0 V_*\approx\frac{4\pi\alpha (GM)^2\rho_0}{(V_*^2+c_s^2)^{3/2}}
\end{equation}
where $\rho_0$ is the ambient gas density. $\alpha$ is a constant of order unity first introduced by \citet{1944MNRAS.104..273B} to account for the non-uniformity of gas velocity inside the Bondi radius due to the gravitational circulation.
\citet{1952MNRAS.112..195B} uses $\alpha=1/2$, but numerical calculations by \citet{1985MNRAS.217..367S} suggest $\alpha=1$.
Due to the long range nature of gravity, gas residing outside the Bondi radius also experiences  gravitational forces.
As a consequence, incoming gas with an impact parameter larger than the Bondi radius is not accreted onto the star, but accumulates behind the star, forming a overdense wake.
Gravity from the overdense wake exerts a drag opposite to the object's direction of motion known as  dynamical friction \citep{1943ApJ....97..255C,1999ApJ...513..252O}.
 For the supersonic case, the acceleration of dynamical friction is given by
\begin{equation} \label{eq:aDF}
    a_{DF}=\frac{4\pi G^2 M\rho_0}{V_*^2}\ln\left[ \Lambda\left(1-\frac{1}{\mathcal{M}^2}\right)^{0.5}\right]
\end{equation}
where $\mathcal{M}=V_*/c_s$ is the Mach number of the object, and $\ln\Lambda=\ln(b_{\max}/b_{\min})$ is the Coulomb logarithm with $b_{\max}$ and $b_{\min}$ being the maximum and minimum impact parameter respectively.
$\ln\Lambda$ quantifies the spatial range over which gravity is important.\footnote{In simulations, we take $b_{\max}$ to be the half box size and $b_{\min}$ to be the smoothing length of gravity.}
Numerical simulations are in generally good agreement with this simple picture \citep{2012ApJ...752...30B}.

The standard analytic model of BHL accretion does not include any outflow from the central object. Outflows, however, are expected to be present in many of the physical situations where BHL accretion occurs. Massive stars can drive winds and previous studies \citep{2016NewAR..75....1S, 2018MNRAS.475.1023M, 2019MNRAS.488.5615S,2019arXiv191204662H} have found that outflow may play an important role in the common envelope evolution of binary stars.
In particular, powerful outflows are observed to accompany super-Eddington accretion on to compact objects in sources such as SS433 \citep{2004ASPRv..12....1F}. 
The influence of such outflows on accretion and dynamical friction remains to be fully explored.
\citet{2019arXiv190601186G} considered the case where a strong isotropic wind is driven from the compact object.
Their analytical calculations showed that when the strong wind shuts off the accretion, a bow shock is launched around the object.
Instead of an overdense wake, an underdense hole is formed behind the bow shock.
The gravitational acceleration of the object from the ambient gas becomes ``negative dynamical friction''\footnote{We refer to a force opposite to the direction of motion as a positive dynamical friction force, and vice versa. Because the star experiences a wind from negative $x$, negative friction forces point along $-x$.} if the wind velocity $V_w$ and the object's velocity $V_*$ satisfy $u\equiv V_*/V_w<1.71$. In addition to this purely mechanical effect, thermal and radiative feedback can also be important \citep{1991ApJ...371..696T,1990ApJ...356..591B,1995A&A...297..739K,2013ApJ...767..163P,2017ApJ...838...13S}. \citet{2017ApJ...838..103P} studied the radiative feedback of accreting massive black holes and found that the overdense wake can be destroyed, reducing the dynamical friction and in some cases reversing its direction. \citet{2017MNRAS.465.3175M} showed that diffusion of heat through an opaque medium surrounding the accreting object similarly leads to a heating force of opposite sign (and in some cases larger magnitude) than the dynamical friction. Related physical effects may be important for the migration of planets (and generically for embedded gravitating objects) in Keplerian disc flows \citep{2017MNRAS.472.4204M}.

In this paper, we carry out a numerical study of a compact object with an outflow moving through uniform ambient gas.
We focus on the case where the compact object is moving supersonically through the gas and study the effects of outflow on the accretion rate and dynamical friction in the case of both a beamed jet as well as an isotropic wind. To make the problem numerically tractable, we parameterize the properties of the outflow and do not self-consistently model the launching mechanism. 
The plan of the paper is as follows.
In Section~\ref{sec:numerical}, we outline our numerical methods and set up for the simulations.
In Section~\ref{sec:isotropic}, we present results for the case of an isotropic wind and compare it to the  analytical calculations of \citet{2019arXiv190601186G}.
The case of a beamed jet is presented in Section~\ref{sec:jet}.
The final section is devoted to discussions.

\section{Numerical Schemes}\label{sec:numerical}

We use the RAMSES code \citep{2002A&A...385..337T} to simulate a compact object moving in a 3D Cartesian box. Ramses is a mature AMR code that includes self-gravity, sink particles \citep{2010MNRAS.409..985D,2014MNRAS.445.4015B}, and radiation \citep{2013MNRAS.436.2188R}.  Previously, one of the authors and his collaborator used RAMSES to study star formation in a turbulent gas both without feedback \citep{2017MNRAS.465.1316M} and with jet feedback \citep{2018MNRAS.475.1023M}.  Here, we use the modifications for sink particles and jet and wind feedback for RAMSES that is described in \citet{2018MNRAS.475.1023M}.  

The sink particle module that we use in RAMSES is the same as described in \citet{2017MNRAS.465.1316M}, but with a few crucial changes.  First, we create the sink particle at the beginning of the simulation, not based on a Jeans criterion as in \citet{2017MNRAS.465.1316M}.  Second, we accrete all the gas above a (low) threshold density within two cells of the sink particle. \footnote{We do not study the effect of varying the accretor size in this paper. Though previous 2D simulations show that flow instabilities will develop for a sufficiently small accretor \citep{1988ApJ...335..862F,2009ApJ...700...95B}, that effect is less significant in 3D simulations \citep{1997A&A...317..793R,1999A&A...346..861R,2012ApJ...752...30B}.} Similarly, we use the same jet and wind feedback module as \citet{2018MNRAS.475.1023M}.  For completeness, we briefly describe the jet and wind feedback module here.  We inject mass, momentum and energy in a region that is $4$ to $8$ cells from the sink particle at the finest level. For the isotropic wind, we use a flat distribution in angles.  For the jet case, we assume a Gaussian jet.  Here, the injection region consists of a bi-cone with a user-defined opening angle of  $90$ degrees ($45^\circ$ half-opening angle) about a chosen jet axis. We note that while the initial angle can be large, recollimation produces a narrower jet on large scales. 

The 3D simulation box is set up in the rest frame of the compact object, where the compact object with mass $M$ is placed at the center of the box, as in \citet{2017ApJ...838...56M}.  
In our simulations, the ambient ideal gas initially has constant density $\rho_0$ flowing along the $+x$ direction with constant velocity $V_*$ (the compact object is moving along $-x$ direction in the lab frame).
The gas obeys an ideal equation of state, with adiabatic index $\gamma = 5/3$.
The boundaries of the computational box are set to be periodic in the $y$ and $z$ direction and ingoing/outgoing for the left/right boundary in the $x$ direction.   
We do not account for deflection of streamlines outside the box, and the box size and resolution are chosen so that $L_B\gg R_B$ to reduce the effects of gravity on the fluid near the boundary.
Our grid spacing $\Delta x$ is chosen to be smaller than $R_B$ to resolve the relevant scale of the accretion process as well as the structure formed by the outflow (e.g. bow shock).
In our typical runs, we use $256$ cells for each side with one level of refinement around the compact object, for an effective resolution of $512^3$. 
We have also performed convergence tests to make sure that our simulations produce consistent results independent of the box size and resolution. 

The compact object is treated as a sink particle in the simulation which accretes ambient gas with its own gravity, which is smoothed over $4$ cells on the finest level.
The simulation includes both the gravity of the compact object and self-gravity of the gas.
However, as we use periodic boundary conditions for our gravity solver, the gravity from the uniform gas background is neglected, i.e. the zeroth Fourier mode of the gas density distribution is ignored.
This affects the self-consistency of our results on the largest scales, as we discuss in \S\,\ref{sec:DynFrict}.
In addition, the compact object is accelerated as it absorbs the momentum of the gas it accretes. (Any outflow emits no net momentum.) 
We update the object's mass and momentum as it accretes, thereby tracking the components of its net acceleration; however, we do not update its position, which is fixed at the box center.  

The relevant parameters and observables in our simulations are the ambient gas density $\rho_0$ and sound speed $c_s$, the velocity $V_*$, mass $M$ and accretion rate $\dot{M}$ of the compact object, the velocity $V_w$ and mass loss rate $\dot{m}_w$ of the outflow.
We will present our results in dimensionless parameters
\begin{equation}
    \mathcal{M}\equiv \frac{V_*}{c_s} \quad\textrm{and}\quad u\equiv \frac{V_*}{V_w}\;.
\end{equation}
The acceleration will be normalized by $a_{DF}$ evaluated with $\Lambda=64$ being the ratio of the half box size over the smoothing length of the gravity.
All other quantities will be normalized by the corresponding values from the Bondi accretion ($\alpha=1$) and we define the Bondi timescale as
\begin{equation}
    t_B\equiv \frac{2GM}{(c_s^2+V_*^2)^{3/2}}.
\end{equation}
We run each simulation for a time much longer than $t_B$ to allow the system to relax.

\section{Isotropic Wind}\label{sec:isotropic}
We first simulate the case where the wind blowing from the star is isotropic with constant mass loss rate $\dot{m}_w$ and velocity $V_w$.
The star is taken to be a $M=1 M_\odot$ compact object and the ambient gas initially has constant density $\rho_0=10^{-6}$~g/cm$^3$ with pressure $p_0$ corresponding to a sound speed of $c_{s0}=38$~km/s.
The fiducial parameters used in the simulation are set as $V_*=600$~km/s and $V_w=3000$~km/s, corresponding to Mach number of $\mathcal{M}\approx 15.8$ and $u=0.2$. 
The Bondi values and dynamical friction for this setup are
\begin{eqnarray}
    R_B &=& 7.4\times 10^{10}\;\mathrm{cm},\nonumber\\
    \dot{M}_B &=& 0.016 \; M_\odot/\mathrm{yr},\nonumber\\
    t_B &=& 1230\;\mathrm{s},\nonumber\\
    a_{DF} &=& 0.15\;\mathrm{cm/s}^2.\nonumber
\end{eqnarray}
We set the box size to be $L_B=10^{13}$~cm, much larger than $R_B$, and adopt the mass loss rate $\dot{m}_w=1 M_\odot$/yr.
Here, $\dot{m}_w$ is much larger than the Bondi accretion rate $\dot{M}_B$ to test the analytical solution given by \citet{2019arXiv190601186G}, which neglected the influence of the object's gravity on the gas flow.

\begin{figure}
    \centering
    \includegraphics[width=.45\textwidth]{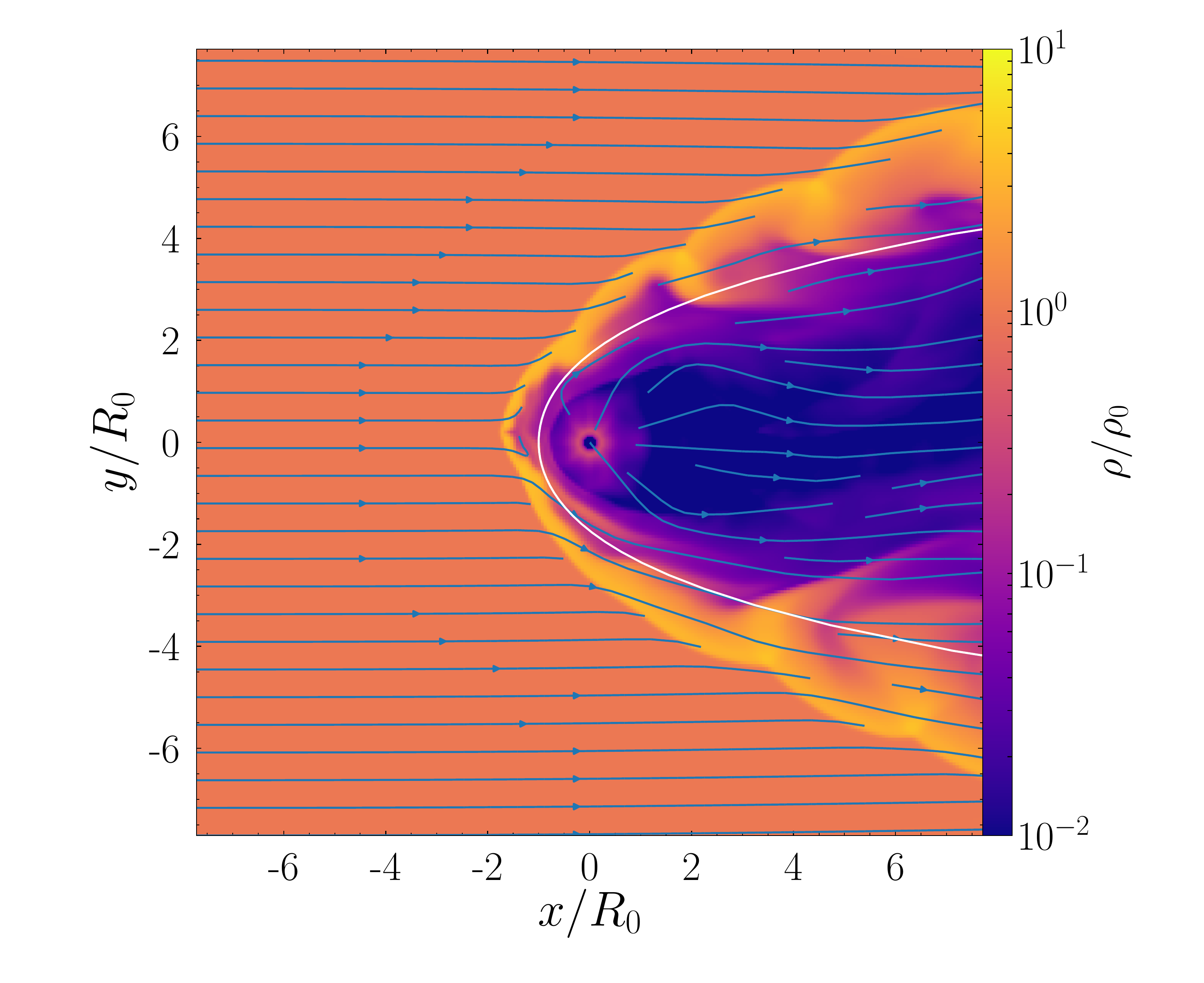}
    \caption{Slice of gas density in the central $x-y$ plane with isotropic wind at $t=630 t_B$. The streamline shows the gas velocity and the white line is the analytical solution from \citet{1996ApJ...459L..31W}.}
    \label{fig:isotropic_slice}
\end{figure}

Figure~\ref{fig:isotropic_slice} shows the gas density slice in the central $x-y$ plane at $t=630 t_B$~.
The color shows the gas density and the blue streamline follows the gas velocity.
The strong wind outflow blocks accretion and redirects the flow, and forms a  high-density bow shock surrounding an under-dense cavity around the compact object. 
At the surface of the bow shock, the gas density exhibits corrugated patterns indicating the presence of the Kelvin-Helmholtz instability.

The white line in Figure~\ref{fig:isotropic_slice} shows the the analytical solution given by \citet{1996ApJ...459L..31W}, which neglects gravitational effects and models the wind-ambient collision as inelastic, resulting in formation of a thin shell. 
The white line marks the position of the shell in the frame of the compact object, given by
\begin{eqnarray}
    R_s = R_0\frac{\sqrt{3(1-\theta\cot\theta)}}{\sin\theta},
\end{eqnarray}
in spherical coordinates for which  the central compact object  is at the origin, and the axis $\theta=0$  is in the negative-$x$ direction. Here $R_0\equiv \sqrt{\dot{m}_w V_w/4\pi \rho_0 V_*^2}$ is the characteristic standoff distance.
The over-density created by the bow shock follows the white line but lies further away from the compact object since the adiabatic ambient shock separates from the contact discontinuity.

\begin{figure}
    \centering
    \includegraphics[width=.45\textwidth]{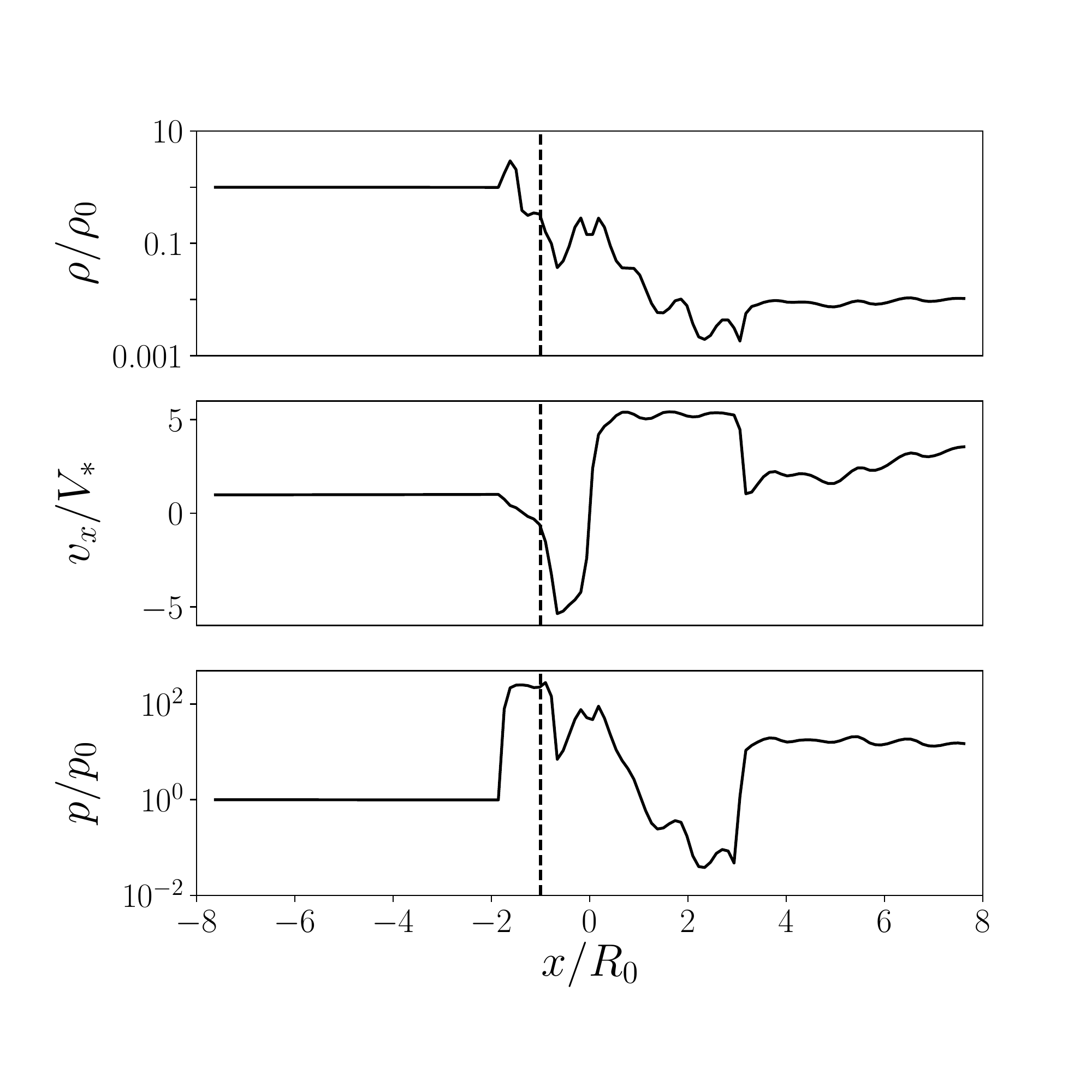}
    \caption{ Gas profile along the $x$-axis at the center. Upper panel: gas density, Middle panel: velocity parallel to the $x$-axis, Bottom panel: pressure. The dashed line is the analytical shock position from \citet{1996ApJ...459L..31W}.}
    \label{fig:isotropic_profile}
\end{figure}

Figure~\ref{fig:isotropic_profile} shows the density, velocity and pressure profiles along the $x$-axis at the center of the computational box.
The dashed line indicates the shock position from the analytical solution by \citet{1996ApJ...459L..31W}. The position of the shock marked by the velocity jump is close to the dashed line, and the density increases ahead of the shock as gas flow from upstream accumulates there.
Moreover, the reverse shock is close to the central compact object, resulting in a large pressure which also pushes the over-density outward.
Behind the compact object, the wind blows away the gas causing the density to drop while the velocity remains constant. 

\citet{2019arXiv190601186G} derive an analytical approximation for the gravitational acceleration on the central compact object, using the matter distribution predicted by \citet{1996ApJ...459L..31W}.
In Wilkin's solution, the density has constant value $\rho_0$ outside the bow shock and drops as $\rho_0 u^2R_0^2/R^2$ from the central compact object inside the bow shock. The surface density of the bow shock follows
\begin{eqnarray}
    \sigma = \rho_0 R_0^2\frac{[2u(1-\cos\theta)+R_s^2\sin^2\theta/R_0^2]^2}{2R_s\sin\theta\sqrt{(\theta-\sin\theta\cos\theta)^2+(R_s^2/R_0^2-1)\sin^4\theta}}.
\end{eqnarray}
The total gravitational force is pointing along the $x$ direction due to symmetry and has contributions from both the gas and the bow shock
\begin{eqnarray}
    F_{\rm gas} &=& - 2\pi GM\int_0^\infty r^2 \md r \int_0^\pi \sin\theta\md\theta \frac{\rho_0}{r^2}\cos\theta, \\
    F_{\rm shock} &=& -2\pi GM \int_0^\pi J_\theta\md\theta R_s\sin\theta \frac{\sigma}{r^2} \cos\theta,
\end{eqnarray}
where $J_\theta$ is the Jacobian
\begin{eqnarray}
    J_\theta\equiv\sqrt{\left(\frac{\partial R_s\sin\theta}{\partial\theta}\right)^2 + \left(\frac{\partial R_s\cos\theta}{\partial\theta}\right)^2}.
\end{eqnarray}
The net force is
\begin{eqnarray}\label{eq:net_force}
    F &=& -\pi GM\rho_0 \int_0^\pi \md\theta \cos\theta \sin\theta R_s \nonumber\\
    &&\times\left[ \frac{3}{2}\left(1+\frac{2u(1-\cos\theta)}{R_s^2\sin^2\theta/R_0^2} \right)^2 - 2\left(1+\frac{u^2}{R_s^2/R_0^2}\right)   \right].
\end{eqnarray}
For $u\ll 1$, the net force asympototes to
\begin{eqnarray}\label{eq:force_asymp}
    F = -8.18 GM\rho_0 R_0
\end{eqnarray}
which points to the same direction as the object's motion.

Equation~\ref{eq:net_force} assumes the ambient gas follows the distribution predicted by \citet{1996ApJ...459L..31W} towards the spatial infinity.
However, in the realistic environment and numerical simulations, the matter distribution deviates from \citet{1996ApJ...459L..31W}'s prediction on sufficiently large scales.  (We discuss this issue further in \S\,\ref{sec:DynFrict}.) We compute the net force due to gas within a sphere of radius $\Tilde{R}$: this can be derived by replacing the upper limit in the radial integral with $\Tilde{R}$.
\begin{eqnarray}
    \int_0^{\Tilde{R}} r^2 \md r \frac{\rho_0}{r^2} &=& \int_0^{R_s} \md r \rho_0 \frac{ u^2 R_0^2}{r^2} + \int_{R_s}^{\Tilde{R}} \md r \rho_0 \nonumber\\
    &=& \rho_0\left(\Tilde{R}-R_s-\frac{u^2 R_0^2}{R_s}\right),
\end{eqnarray}
for $\theta<\Tilde{\theta}$ where the bow shock is inside the sphere. 
The angle $\Tilde{\theta}$ is defined to be where the bow shock crosses the sphere $R_s(\Tilde{\theta})=\Tilde{R}$.
For the range of polar angle where the bow shock is outside the sphere
\begin{eqnarray}
    \int_0^{\Tilde{R}} r^2 \md r \frac{\rho_0}{r^2} = \int_0^{\Tilde{R}} \md r \rho_0\frac{ u^2 R_0^2}{r^2} = -\rho_0\frac{u^2 R_0^2}{\Tilde{R}}.
\end{eqnarray}
The upper limit in the force from the shock should be be replaced by $\Tilde{\theta}$ in the angular integral since the bow shock outside the sphere will not contribute to the gravity.
Therefore, the expression for the net gravity of gas within a sphere of radius $\tilde{R}$ is
\begin{eqnarray}\label{eq:force_finitesph}
    F(\Tilde{R}) &=& -\pi GM\rho_0 \int_0^{\Tilde{\theta}} \md\theta \cos\theta \sin\theta R_s \nonumber\\
    &&\times\left[ \frac{3}{2}\left(1+\frac{2u(1-\cos\theta)}{R_s^2\sin^2\theta/R_0^2} \right)^2 - 2\left(1+\frac{u^2}{R_s^2/R_0^2}\right)   \right] \nonumber \\
    &&- \pi GM\rho_0\left( \Tilde{R}+\frac{u^2 R_0^2}{\Tilde{R}^2}\right) \sin^2\Tilde{\theta} .
\end{eqnarray}

Equation~\ref{eq:force_finitesph} asymptotes to Equation~\ref{eq:net_force} for large $\Tilde{R}$ at the order of $\mathcal{O}(\Tilde{R}^{-1})$.
\begin{figure}
    \centering
    \includegraphics[width=.45\textwidth]{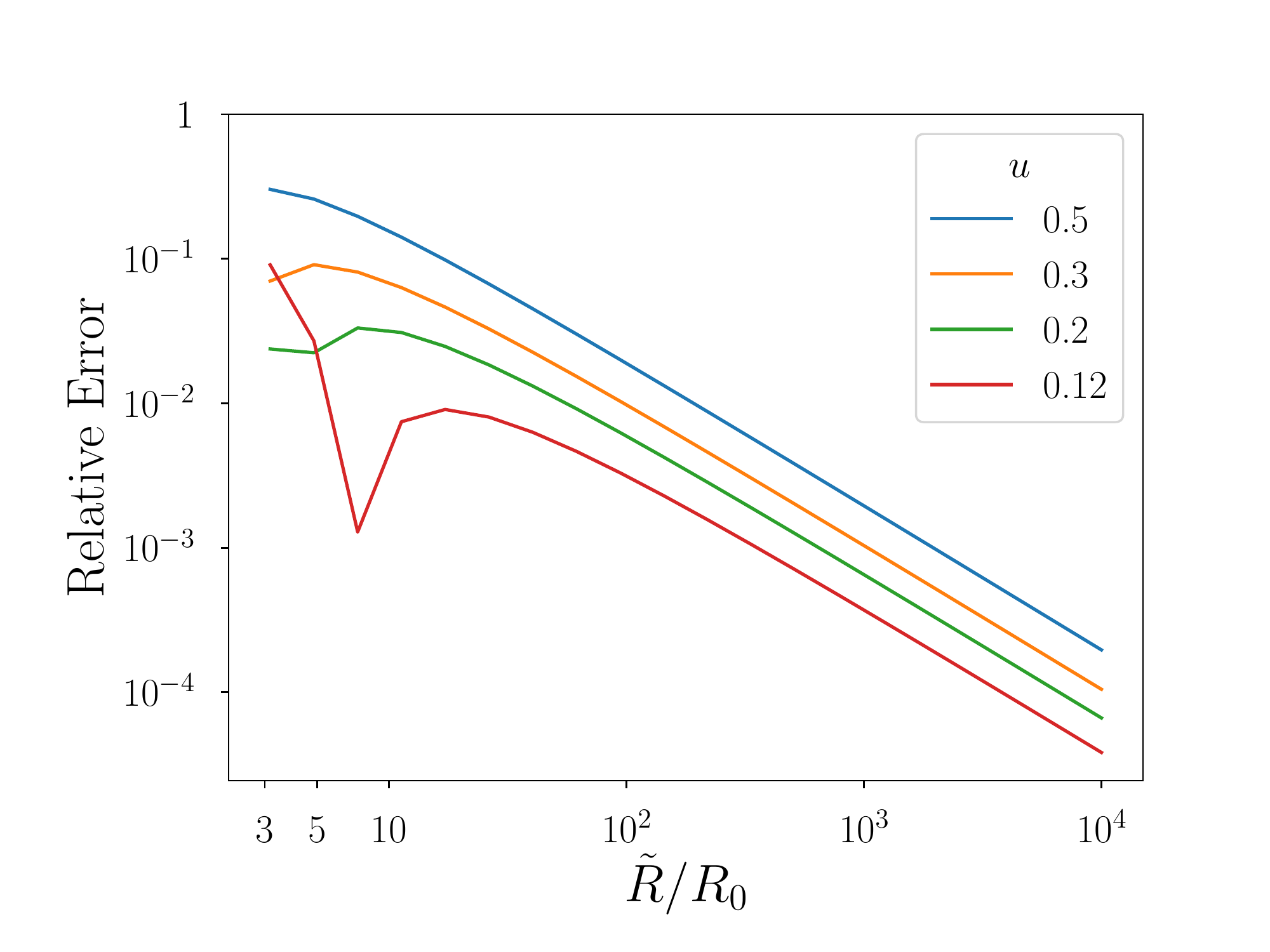}
    \caption{Relative error of Equation~\ref{eq:force_finitesph} as a function $\Tilde{R}$ for various $u\equiv V_*/V_w$.}
    \label{fig:finite_converge}
\end{figure}
Figure~\ref{fig:finite_converge} shows the relative error of Equation~\ref{eq:force_finitesph} as a function of $\Tilde{R}$ for various $u$, and shows this $R^{-1}$ scaling. 
For $u<0.5$ and $\Tilde{R}\gtrsim 5 R_0$, Equation~\ref{eq:force_finitesph} agrees with Equation~\ref{eq:net_force} for better than $10\%$.
On this basis, we would expect the net gravitational acceleration obtained from the simulation to be close to the value given be Equation~\ref{eq:net_force} as long as our computational box is larger than $\sim 5 R_0$ for $u<0.5$.
For the fiducial parameters we use, we have $L_B/2R_0=7.7$, and we should expect convergence to better than $10\%$.\footnote{  
However, we note that Equation~\ref{eq:net_force} is based on a mass distribution derived by neglecting gravity; on sufficiently large scales we should expect the numerically-derived force to eventually deviate from it  -- see \S\,\ref{sec:DynFrict}.}

\begin{figure}
    \centering
    \includegraphics[width=.45\textwidth]{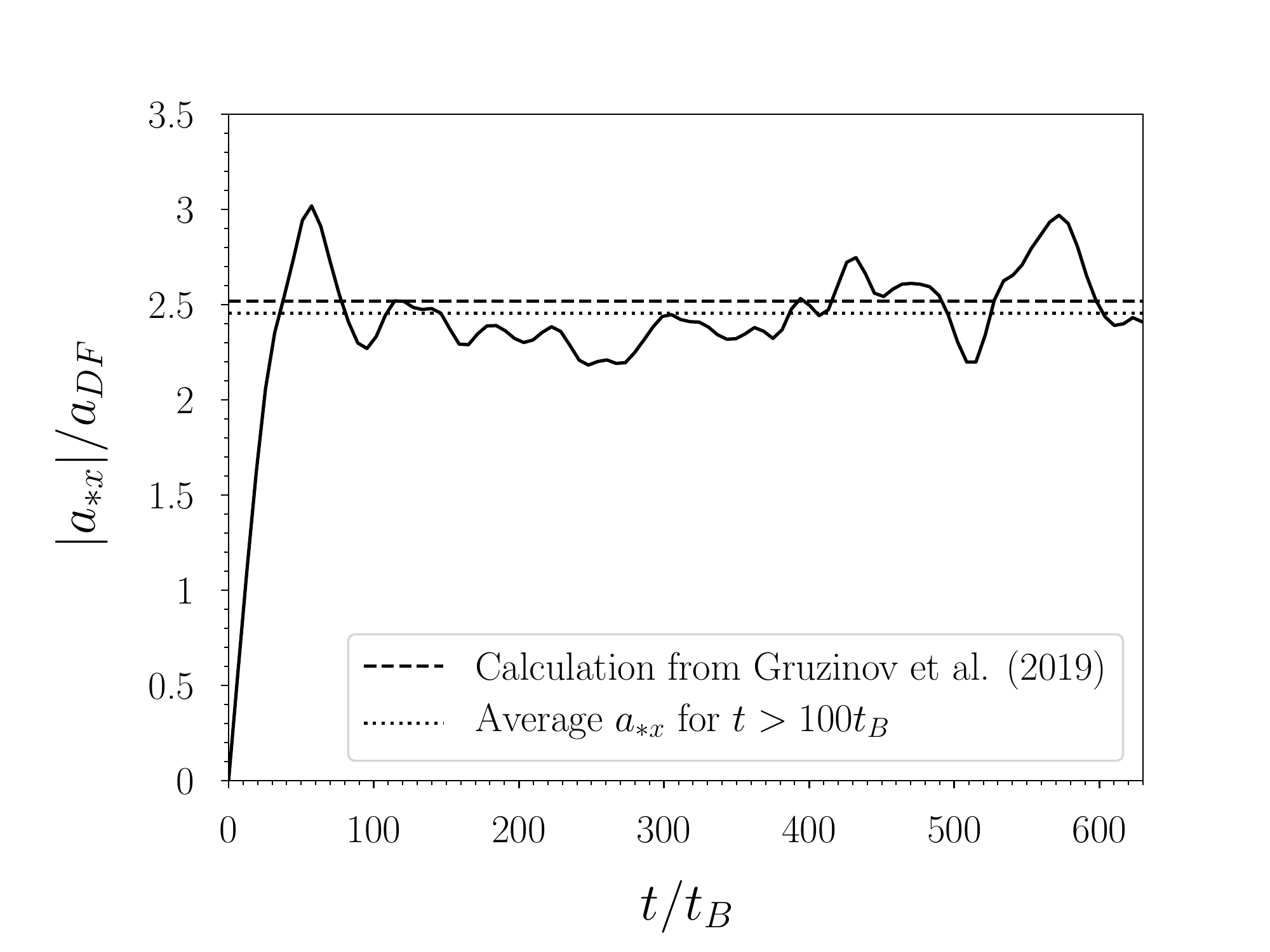}
    \caption{Evolution for the magnitude of the gravitational acceleration of the compact object in the $x$ direction from the ambient gas. The acceleration points in the $-x$ direction and has negative values.
    The dotted line is the average acceleration for $t>100t_B$ and the dashed line is the theoretical calculation from \citet{2019arXiv190601186G}. }
    \label{fig:isotropic_acceleration}
\end{figure}

Figure~\ref{fig:isotropic_acceleration} shows the evolution of the gravitational acceleration for the compact object in the $x$ direction from the ambient gas. After an initial stage of increase where the wind flows out to form a steady bow shock, the acceleration of the compact object settles to oscillate around a steady value.
The dotted line is the average acceleration for $t>100t_B$~years and the dashed line is the theoretical calculation from \citet{2019arXiv190601186G}.
We can see that the two values are close to each other.
The value from \citet{2019arXiv190601186G} is larger since their calculation assumes the over-density sits at the shock, but our simulation suggests that the over-density is farther away.
We have also compared the acceleration obtained with high resolution simulations and larger box size, with entirely consistent results.

\begin{figure}
    \centering
    \includegraphics[width=.45\textwidth]{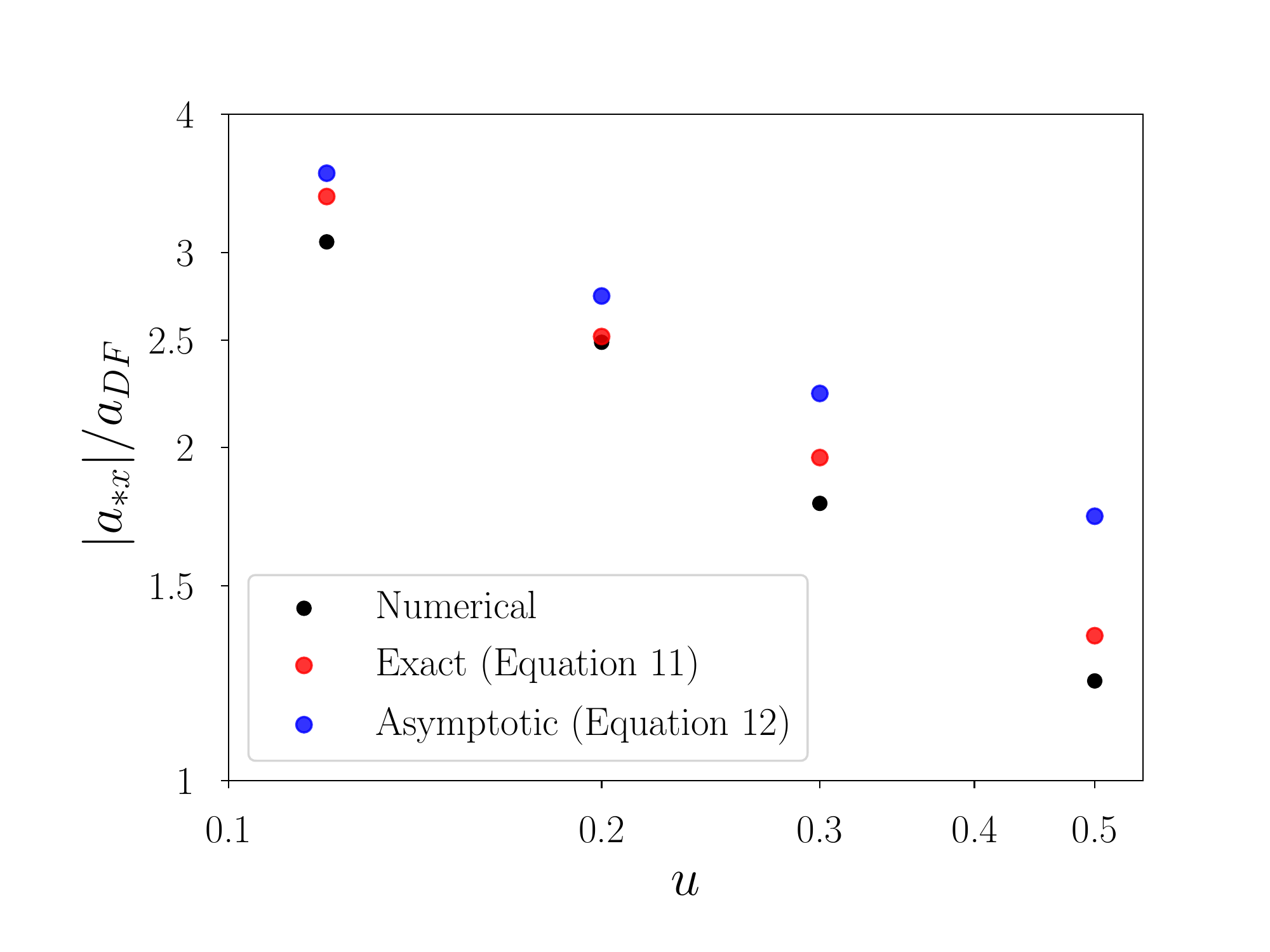}
    \caption{The dependence of acceleration on the wind speed $V_w$. Black dots are simulation results. Red dots are analytical results (Equation~\ref{eq:net_force}) and blue dots are the asymptotic expression for $u\ll 1$ (Equation~\ref{eq:force_asymp}).}
    \label{fig:ax_vw}
\end{figure}

Figure~\ref{fig:ax_vw} shows the gravitational acceleration for different values of $u$ (varying  $V_w$ while holding $V_*$ constant) with all other parameters kept fixed.
The ratio $L_B/2R_0=17.24\sqrt{u}>5$ for all the cases considered in the plots.
Black dots are simulation results which is the average acceleration after a steady state is reached.
Red dots are analytical calculations (Equation~\ref{eq:net_force}) and blue dots are the asymptotic expression for $u\ll 1$ (Equation~\ref{eq:force_asymp}).
We can see that the simulation results are all within $\sim10\%$  of the analytical results, and that the asymptotic formula for $u\ll 1$ tends to overestimate the acceleration for non-zero $u$.

\section{Accretion-Powered Jet}\label{sec:jet}
The numerical results presented in Section~\ref{sec:isotropic} is for the purpose of testing the analytical solution, where we use a fixed mass loss rate $\dot{M}$ larger than the Bondi accretion rate.
In this section, we study a more physical scenario where a compact object moves on the background of ambient gas with accretion powered outflow.
The compact object still has mass of $1M_\odot$, but launches beamed jet rather than isotropic wind.
We set the star velocity to be $V_*=300$~km/s corresponding to Mach number $\mathcal{M}\approx 7.9$.
The Bondi values and the predicted dynamical friction force for this setup are
\begin{eqnarray}
    R_B &=& 3 \times 10^{11}\;\mathrm{cm},\nonumber\\
    \dot{M}_B &=& 0.13 \; M_\odot/\mathrm{yr},\nonumber\\
    t_B &=& 9735\;\mathrm{s},\nonumber\\
    a_{DF} &=& 0.6\;\mathrm{cm/s}^2.\nonumber
\end{eqnarray}
Again, we set the box size to be $10^{13}$~cm, much larger than $R_B$.

\subsection{No Outflow}

\begin{figure}
    \centering
    \includegraphics[width=.45\textwidth]{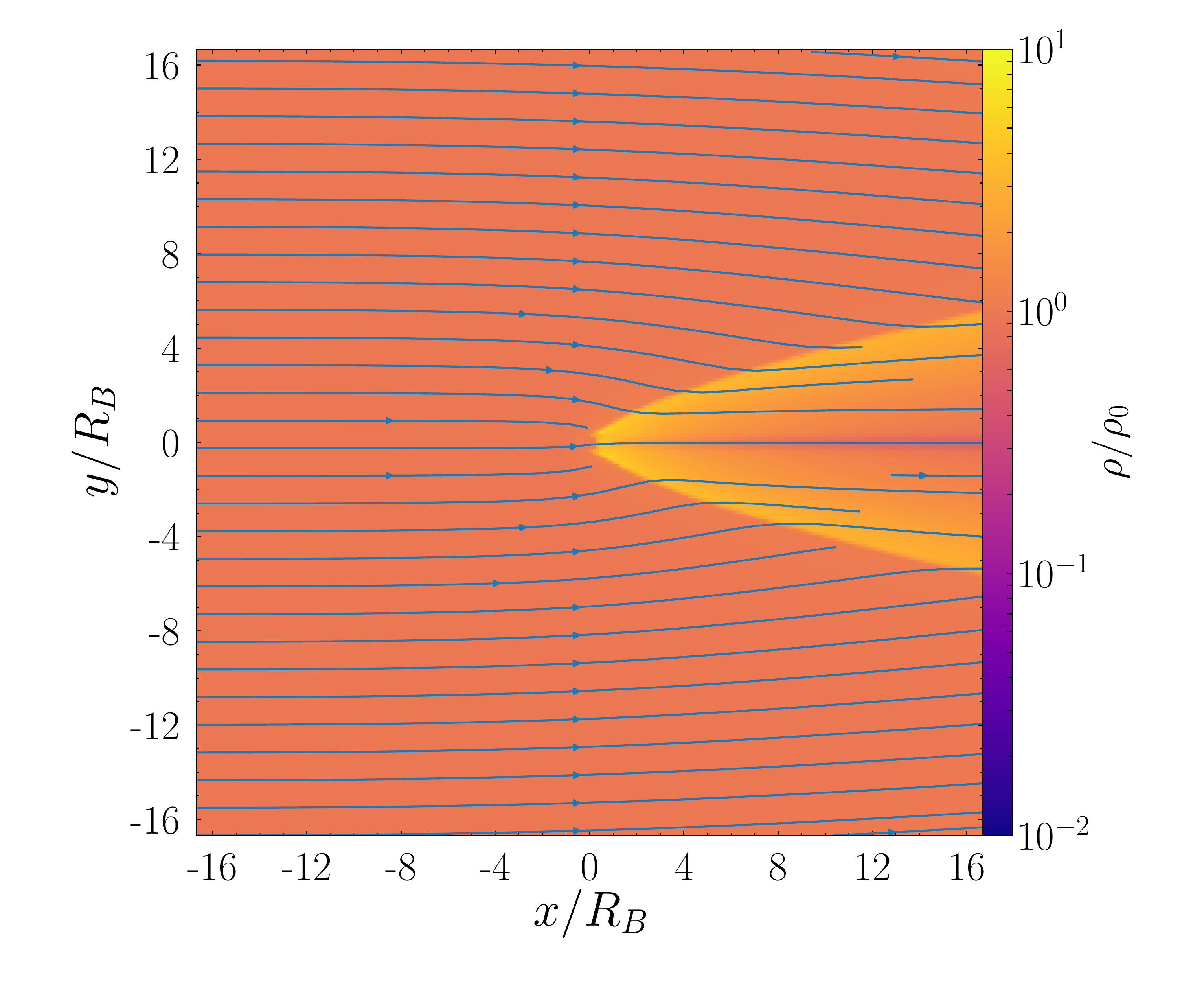}
    \caption{Slice of gas density in the $x-y$ plane for the simulation with no outflow at $t=80t_B$.}
    \label{fig:nojet}
\end{figure}

As a benchmark for comparison, we first run simulations with no outflow.
The supersonic motion of the object leaves an axis-symmetric conic wake behind.
Figure~\ref{fig:nojet} shows the slice of gas density in the $x-y$ plane.
The velocity streamline of ambient gas is refracted towards the object by the gravitational drag until it reaches the wake.
The overdense gas inside the wake acts gravitationally to decelerate the object which is the standard dynamical friction.  
The values of mass accretion rate and dynamical acceleration are consistent with conventional analytical values given by Equation~\ref{eq:acc_rate} and ~\ref{eq:aDF} as shown in Figure~\ref{fig:mdot_jetz} and ~\ref{fig:ax_jetz} for the case of no outflow.

\subsection{Beamed Jet}
For accretion powered outflow, the mass outflow rate is expected to be a fraction of the mass accretion rate.
Therefore, we set the mass loss rate $\dot{m}_w$ to be a constant fraction 
\begin{equation}
    f\equiv \frac{\dot{m}_w}{\dot{M}}
\end{equation} 
of the numerical mass accretion rate $\dot{M}$.
As discussed in Section~\ref{sec:numerical}, $\dot{M}$ is calculated by accreting all the gas within two cells from the sink particle at the finest level.
The shape of the outflow has a Gaussian profile spreading over a polar angle $\theta=45^\circ$ from the beam center.
We vary the parameter $u$ by changing the outflow velocity $V_w$ while keeping $V_*$ constant and make sure that the size of the consequent bow shock remains smaller than the box size of $L_B=10^{13}$~cm.

\begin{figure}
    \centering
    \includegraphics[width=.42\textwidth]{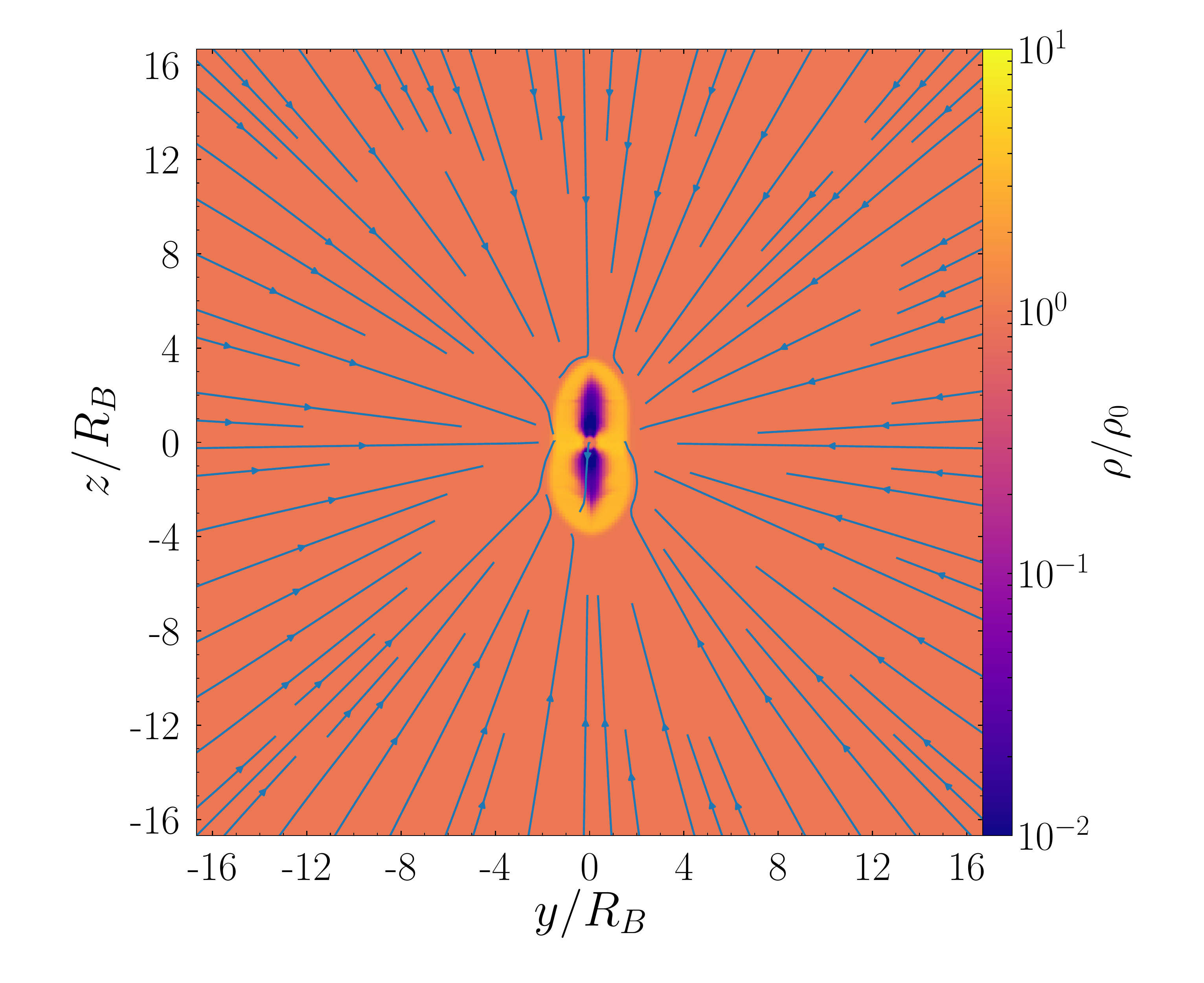}
    \includegraphics[width=.42\textwidth]{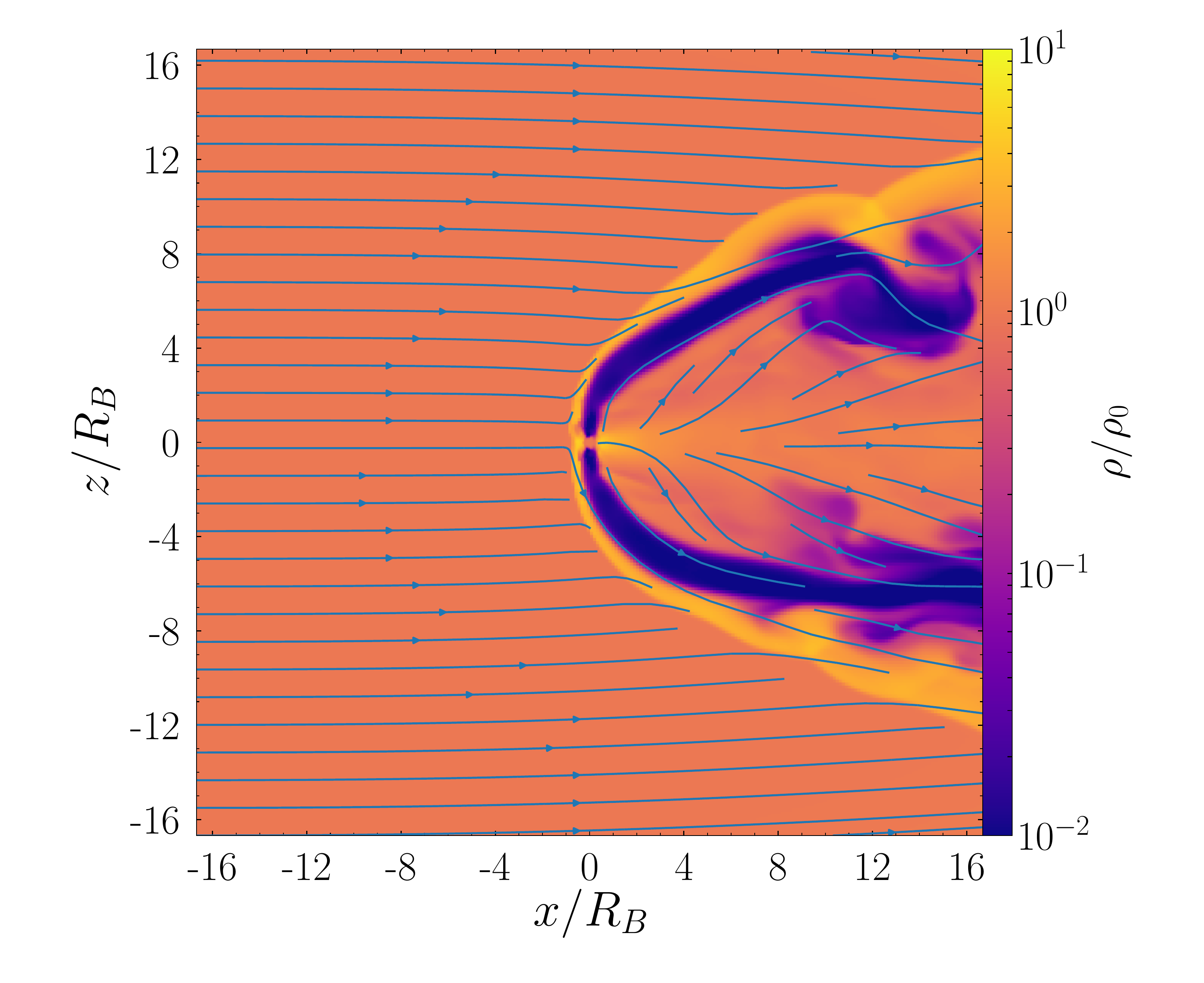}
    \includegraphics[width=.42\textwidth]{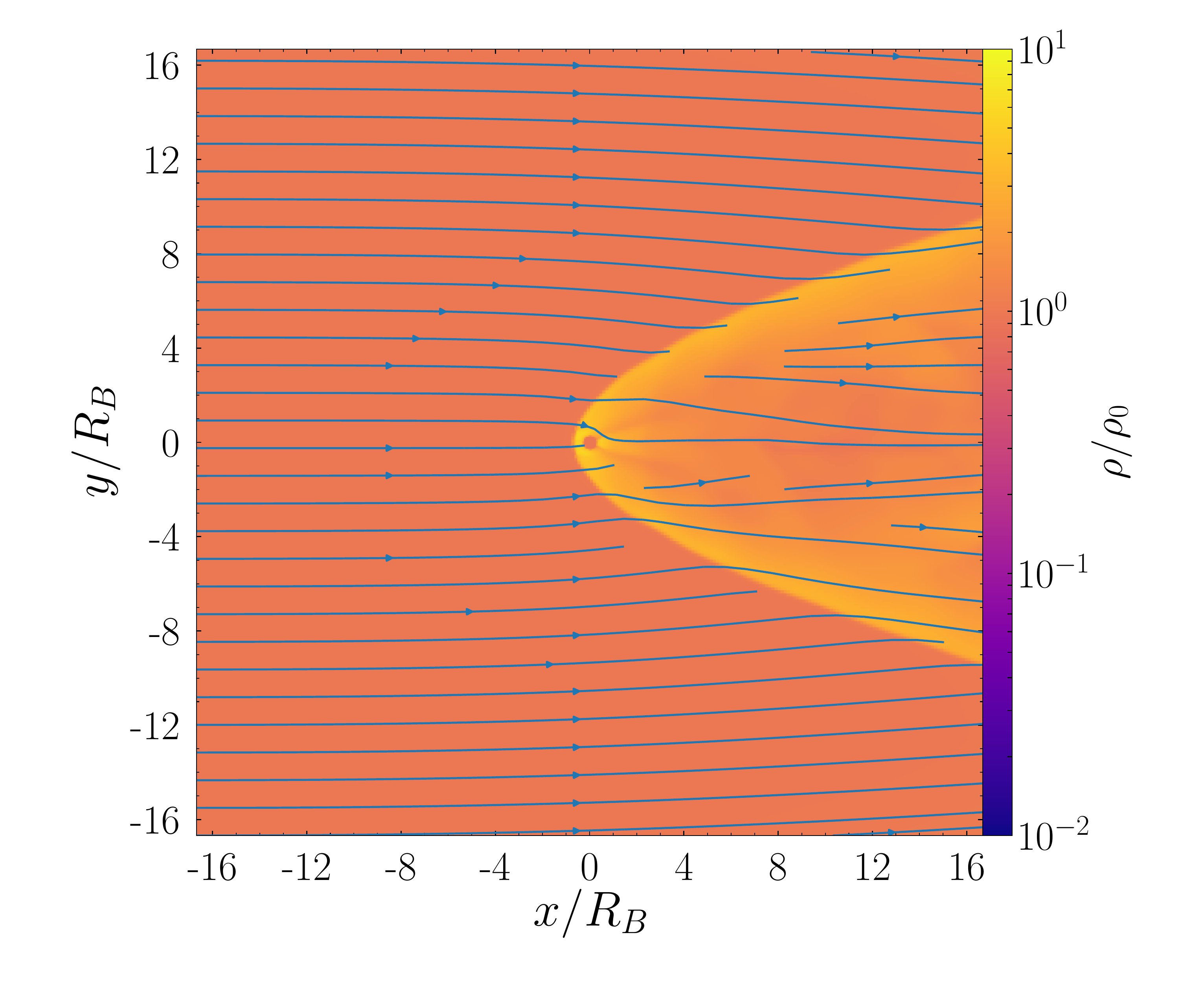}
    \caption{Slices of gas density for three planes passing through the computation box center at $t=80t_B$ after the simulation starts. The outflow is beamed around the $z$-axis perpendicular to the velocity of the compact object with $u=0.03$ ($V_w=10^4$~km/s) and $f=0.3$.}
    \label{fig:jetz}
\end{figure}

Figure~\ref{fig:jetz} shows the slices of the simulation where the outflow is beamed around the $z$-axis (at right angles to the incoming gas) with $u=0.03$ ($V_w=10^4$~km/s) and $f=0.3$.
The slices are plotted in three planes passing through the box center.
In the $y-z$ slice on the top, a bipolar outflow is observed.
A shock front with gas over-density is formed $\sim4R_B$ in the $z$ direction from the central object.
Close to the compact object, there is a under dense gas bubble along the $z$-axis, but over dense gas along the $y$-axis.
In the $x-z$ slice at the center, a bow shock is clearly seen as the beamed outflow encounters the ambient gas.
The jet sweeps up the gas, creating an underdense tunnel which extends along the $z$-axis close to the compact object but is bent  in the $x$-direction by the ambient gas far away from the object.
Compared to Figure~\ref{fig:isotropic_slice}, the under-density inside the bow shock less prominent, for two reasons: the outflow is now beamed, so it sweeps a smaller region; and the mass outflow rate is much smaller.
Moreover, there is an overdense region between the underdense tunnels swept out by the jet.
The $x-y$ slice on the bottom is similar to Figure~\ref{fig:nojet} without outflow, indicating gas not swept by the jet still accumulates behind the star due to  gravitational drag.

\begin{figure}
    \centering
    \includegraphics[width=.45\textwidth]{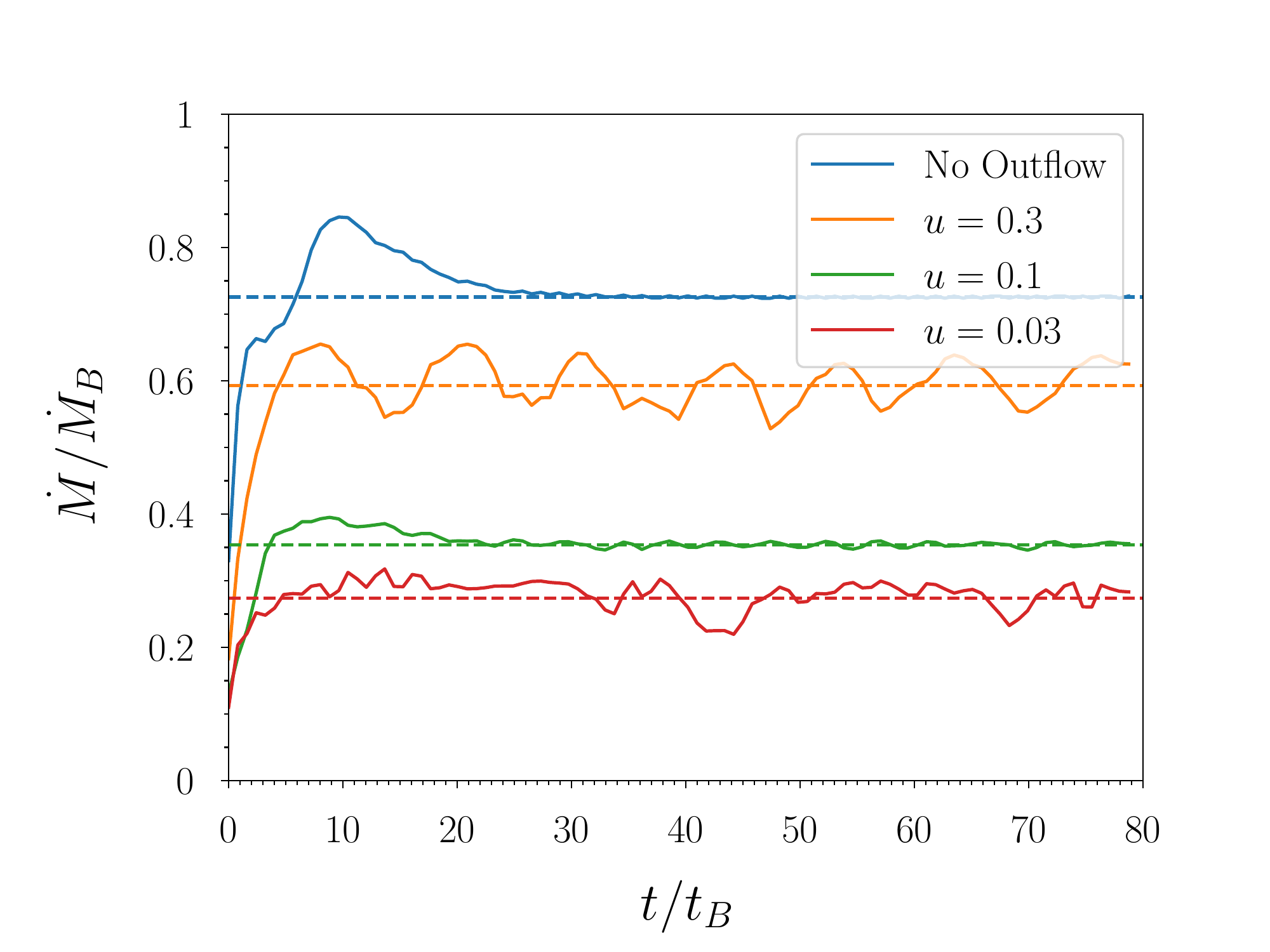}
    \caption{Comparison of mass accretion rate evolution for different simulations. The beamed jet is launched along $z$-axis with $f=0.3$ and various with various $u$ (various $V_w$, $V_*$ is kept constant). The dashed lines show the average acceleration for $t>30t_B$.}
    \label{fig:mdot_jetz}
\end{figure}

Figure~\ref{fig:mdot_jetz} shows the comparison of accretion rate evolution between simulation with and without the outflow.
The mass of compact object increases as it accretes matter from the gas, and therefore the accretion rate should also change as $M^2$ in the simulation. 
However, since the accretion rate is low ($~0.1$~$M_\odot$\,yr$^{-1}$) and the simulation time is short, the change of mass is negligible ($\sim 10^{-3}$).
Therefore, the accretion rate appears to be stationary after an initial phase of relaxation. 
The measured accretion rate for the case without outflow is $\dot{M}_0=0.7$~$\dot{M}_B$, consistent with the BHL accretion theory.
As the jet is turned on, the outflow disrupts inflowing gas and reduces the accretion rate onto the object.
The mass accretion rate keeps dropping with the increasing outflow velocity of the gas.

\begin{figure}
    \centering
    \includegraphics[width=.45\textwidth]{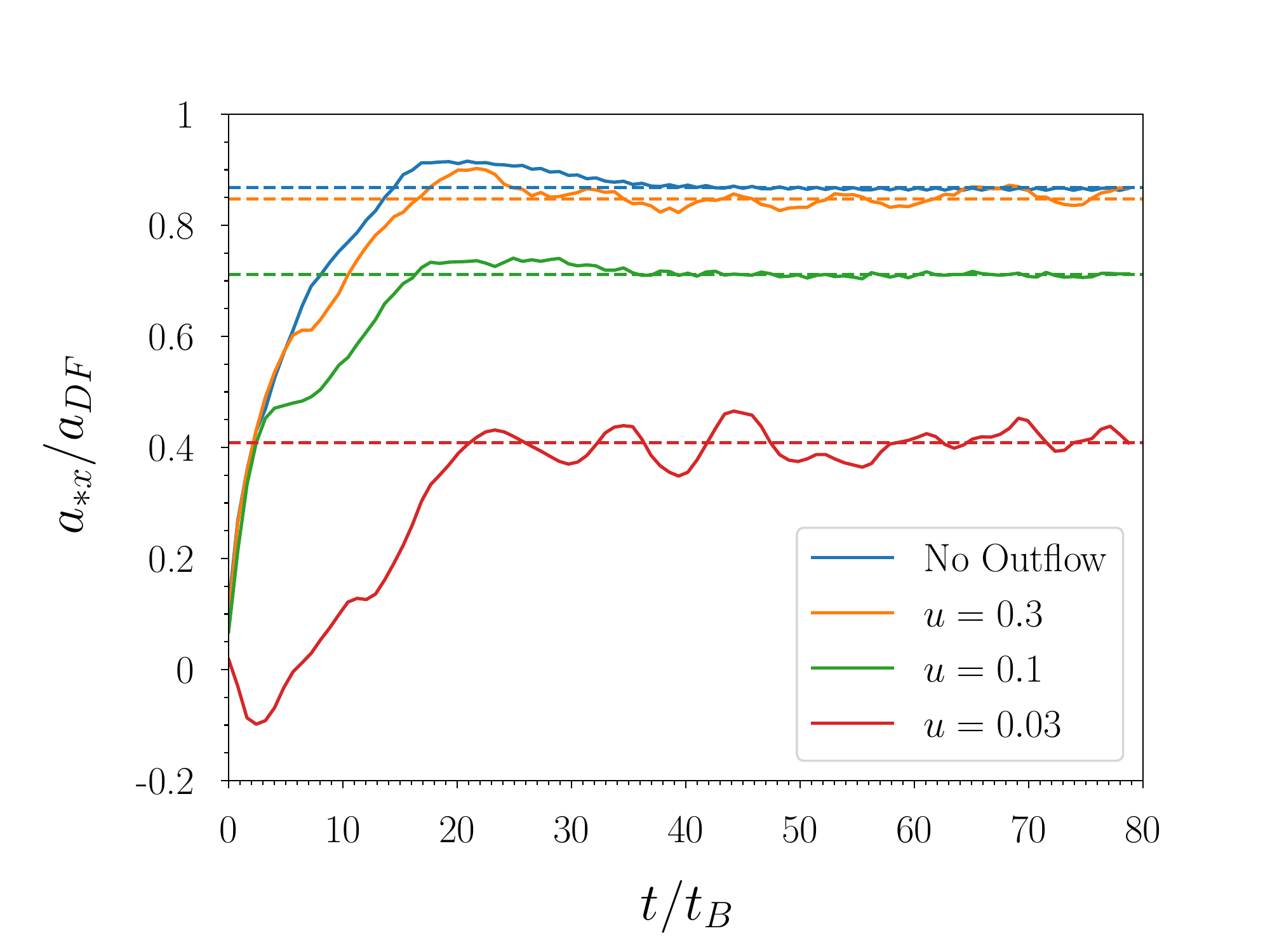}
    \caption{Comparison of acceleration of the object in the $x$-direction for different simulations. The beamed jet is launched along $z$-axis with various $u$ (various $V_w$, $V_*$ is kept constant). The dashed lines show the average acceleration for $t>30t_B$.}
    \label{fig:ax_jetz}
\end{figure}

Figure~\ref{fig:ax_jetz} shows the net acceleration of the star along its direction of motion.
The acceleration is dominated by the gravitational force from the ambient gas, but also includes contributions ($\sim 10\%$) from the accreted momentum.
Without the outflow, the gas accumulated behind the star acts as the source of the standard dynamical friction, and its magnitude is close to our estimation of $a_{DF}$.
As the outflow velocity $V_w$ increases, the dynamical friction is reduced by the underdense tunnels created by the jet.
However, no stable negative dynamical friction is observed in our simulations.
Only a transient stage of negative acceleration is observed during the early epoch for the case with strongest outflow $u=0.03$ ($V_w=10^4$~km/s).

\begin{figure}
    \centering
    \includegraphics[width=.42\textwidth]{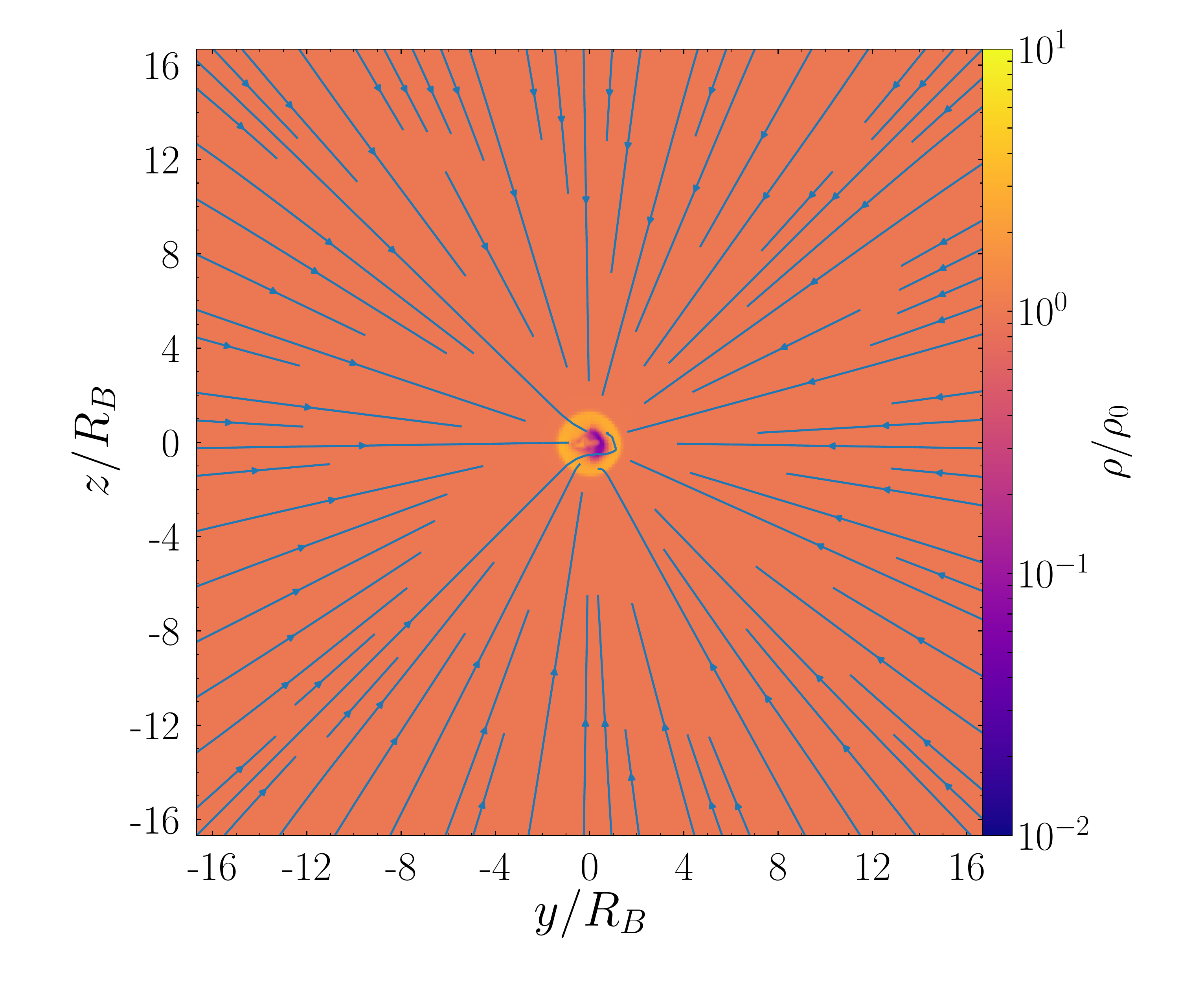}
    \includegraphics[width=.42\textwidth]{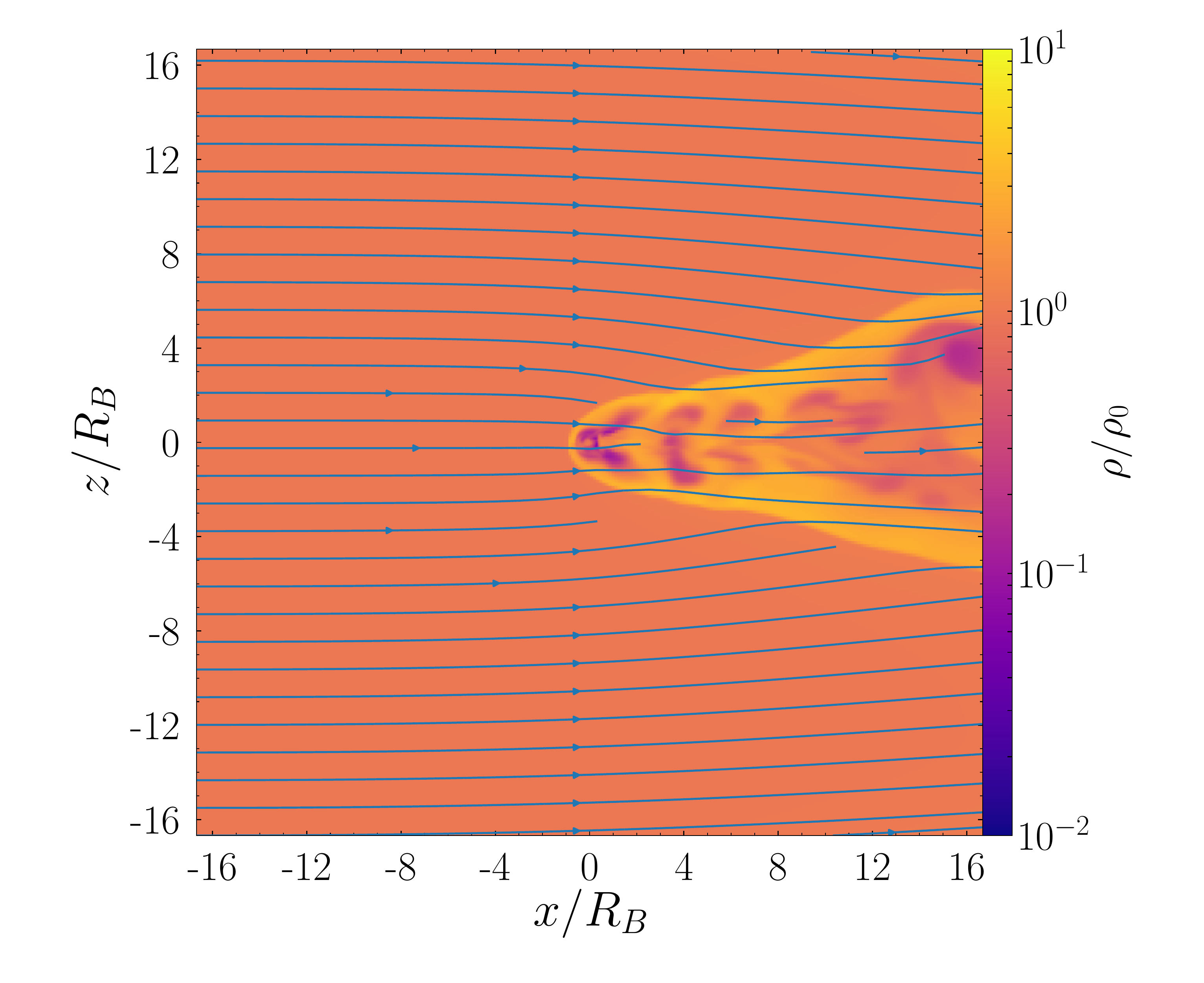}
    \includegraphics[width=.42\textwidth]{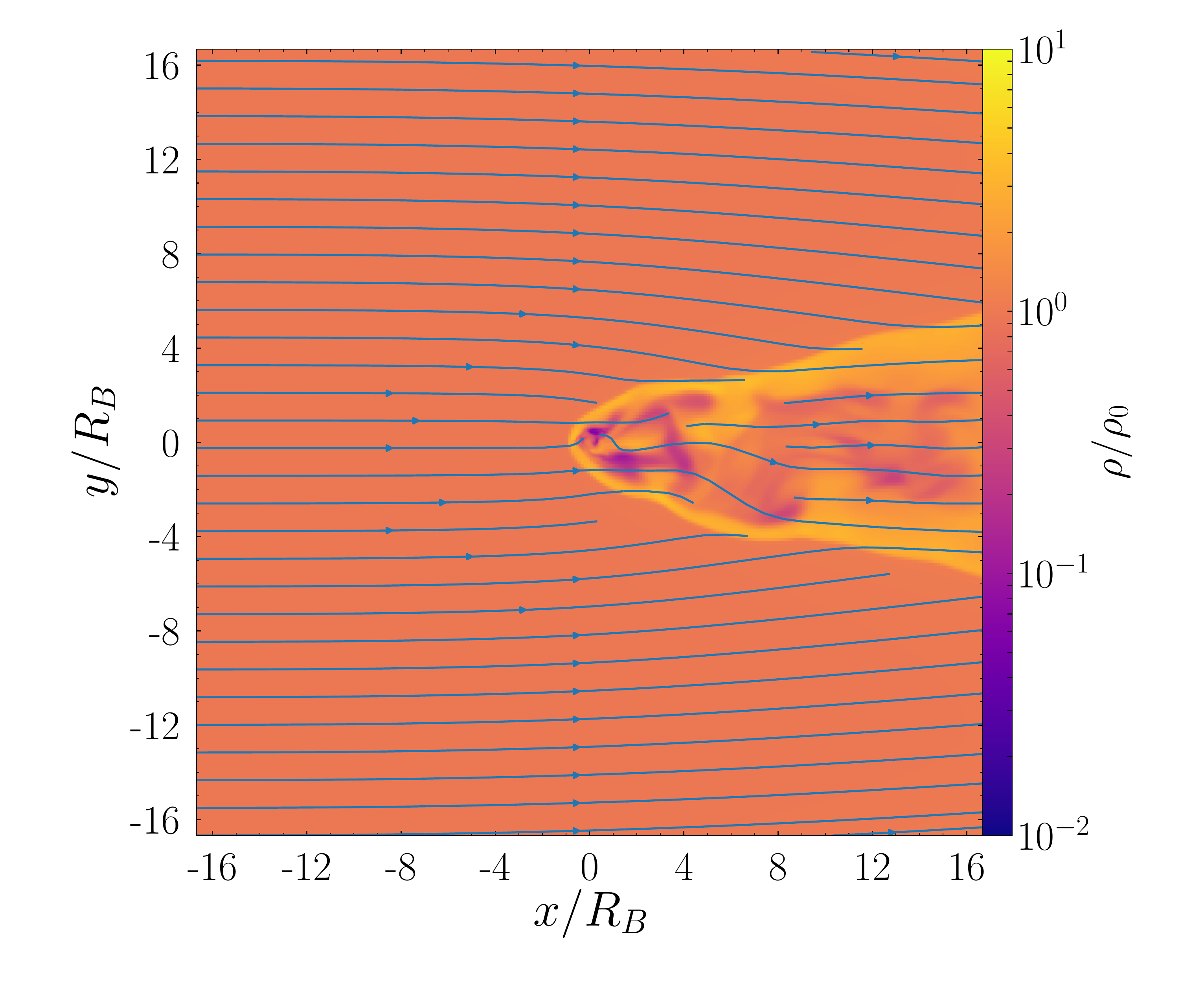}
    \caption{Slices of gas density for three planes passing through the computation box center at $t=80 t_B$~. The outflow is beamed around the $x$-axis perpendicular to the velocity of the compact object with $u=0.03$ ($V_w=10^4$~km/s) and $f=0.3$.}
    \label{fig:jetx}
\end{figure}

In addition to the scenario in which the jet launched perpendicular to the direction of the object's motion, we have also simulated cases where the jet is parallel. This could be relevant if its orientation aligns with the spin of the compact object, which could point in any direction.
Figure~\ref{fig:jetx} shows slices for the parallel case with $u=0.03$ ($V_w=10^4$~km/s) and $f=0.3$.
There is a region of under density around the compact object seen in all three plots.
Unlike the perpendicular case, there is no tunnel of under-density observed.

\begin{figure}
    \centering
    \includegraphics[width=.45\textwidth]{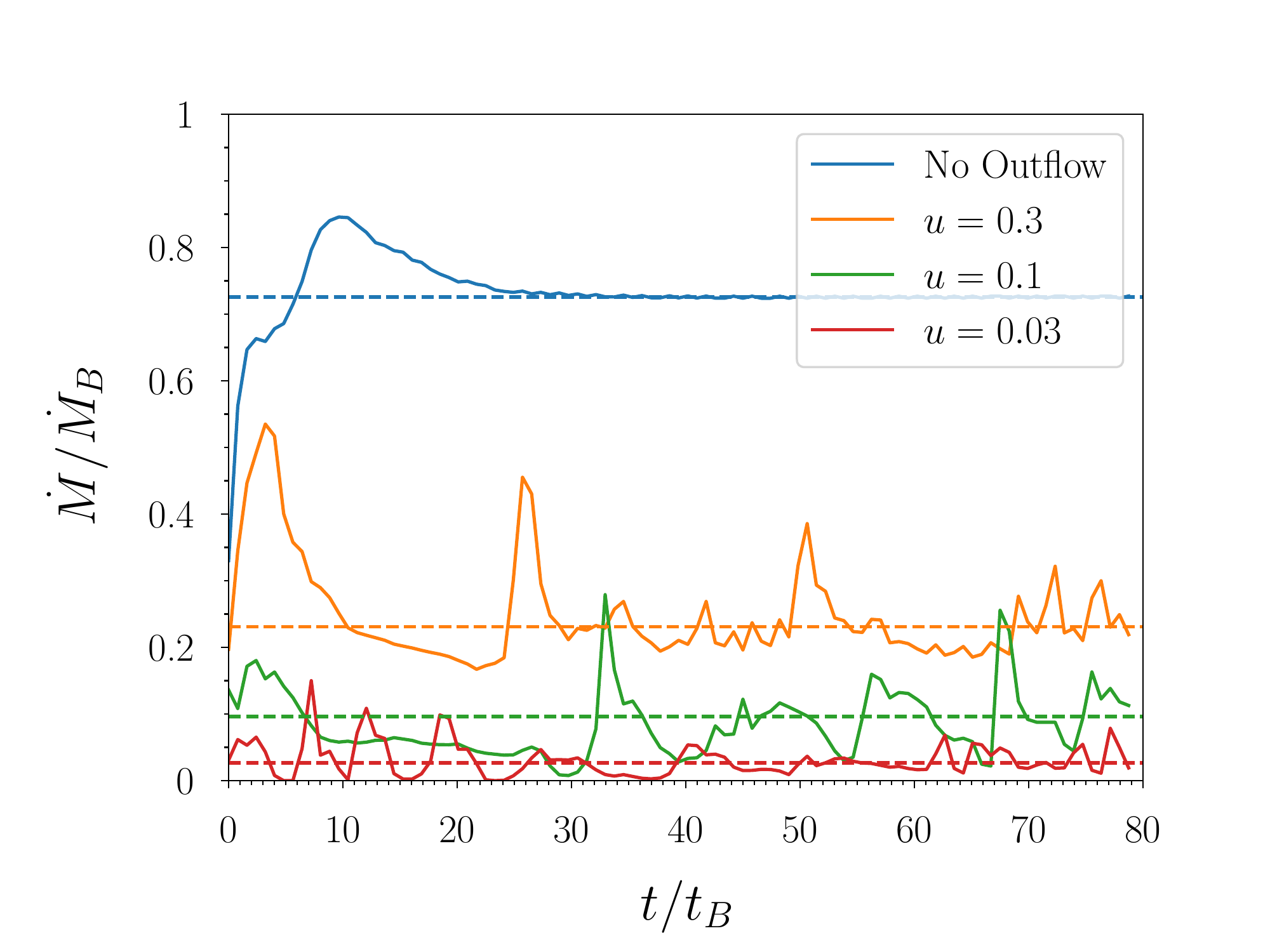}
    \caption{Comparison of mass accretion rate evolution for different simulations. The beamed jet is launched along $x$-axis with $f=0.3$ and various $u$ (various $V_w$, $V_*$ is kept constant). The dashed lines show the average acceleration for $t>30t_B$.}
    \label{fig:mdot_jetx}
\end{figure}

Figure~\ref{fig:mdot_jetx} shows the accretion rate evolution when the outflow is beamed in the $x$ direction (along the flow axis).
Unlike a perpendicular jet, there is no steady state in this case: sporadic spikes in $\dot{M}$ punctuate a low level of accretion.
When the accretion rate is high, the mass loss rate also increases which feeds back negatively on the accretion rate that in turn reduces the mass loss rate.
As the outflow suffers from the head-on collision with the ambient gas, the mixing of streams can excite gas instabilities and break the periodicity of the evolution leading to intermittent spikes.
Signatures of the spikes in accretion rate are also present in the slice plots, where sporadic low density patterns are ejected from the object which then propagate away in the overdense wake.

\begin{figure}
    \centering
    \includegraphics[width=.45\textwidth]{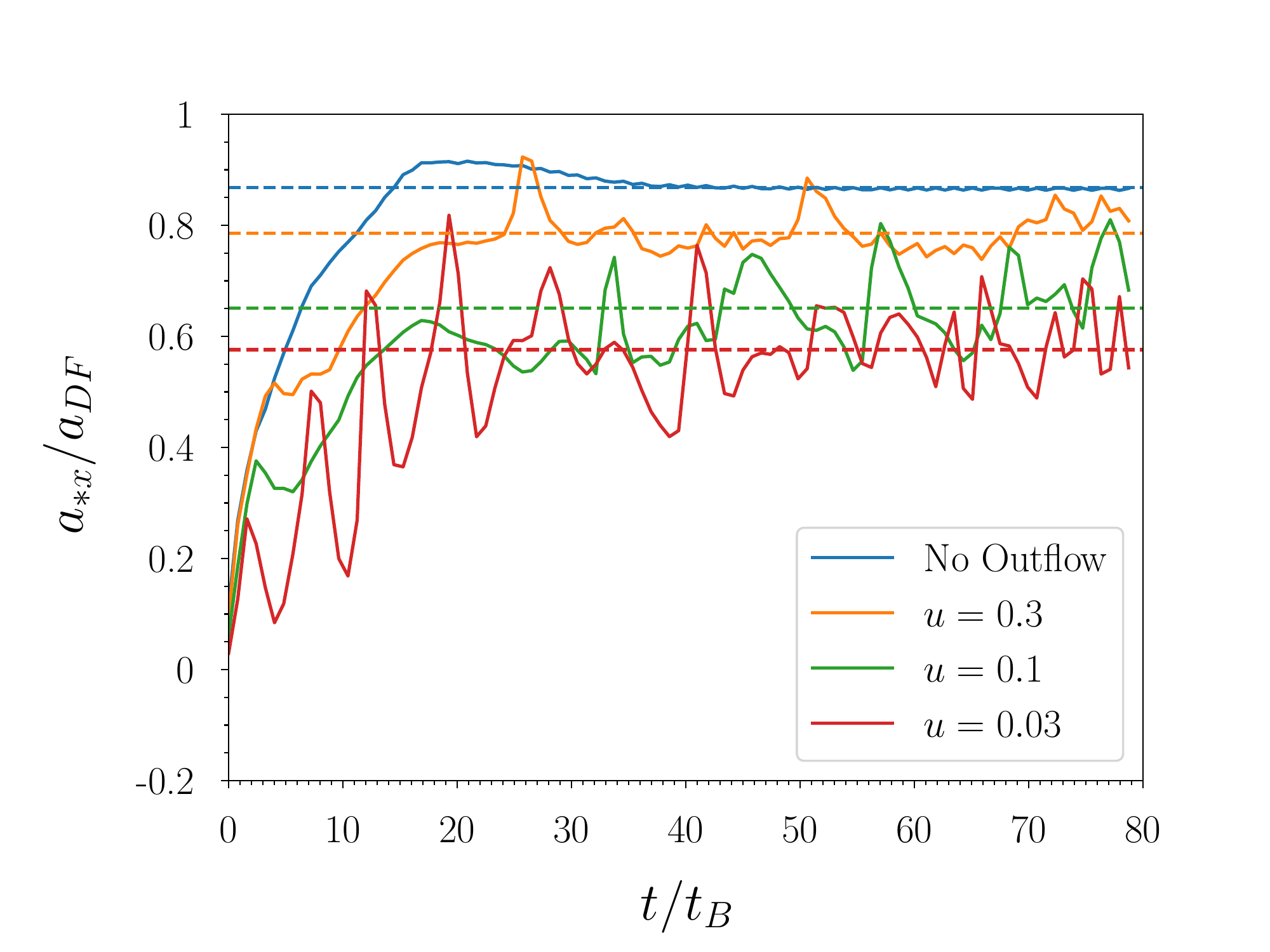}
    \caption{Comparison of acceleration of the object in the $x$-direction for different simulations. The beamed jet is launched along $x$-axis with various $u$ (various $V_w$, $V_*$ is kept constant). The dashed lines show the average acceleration for $t>30t_B$.}
    \label{fig:ax_jetx}
\end{figure}
The intermittent pattern is also present in the acceleration of the object as shown in Figure~\ref{fig:ax_jetx}.
Even though the acceleration is more irregular, its values are still positive with no negative dynamical friction observed in the simulations.

\begin{figure}
    \centering
    \includegraphics[width=.45\textwidth]{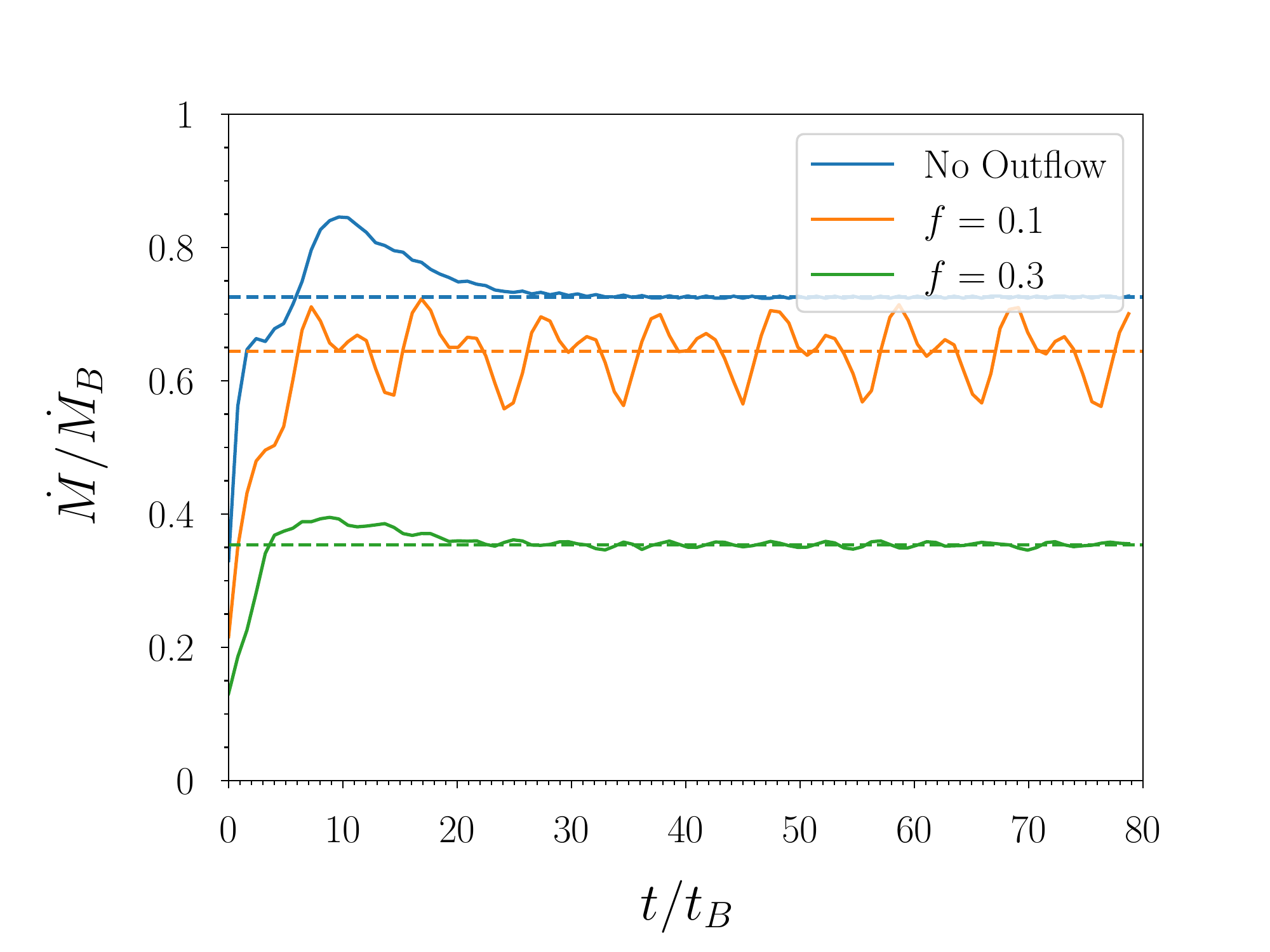}
    \includegraphics[width=.45\textwidth]{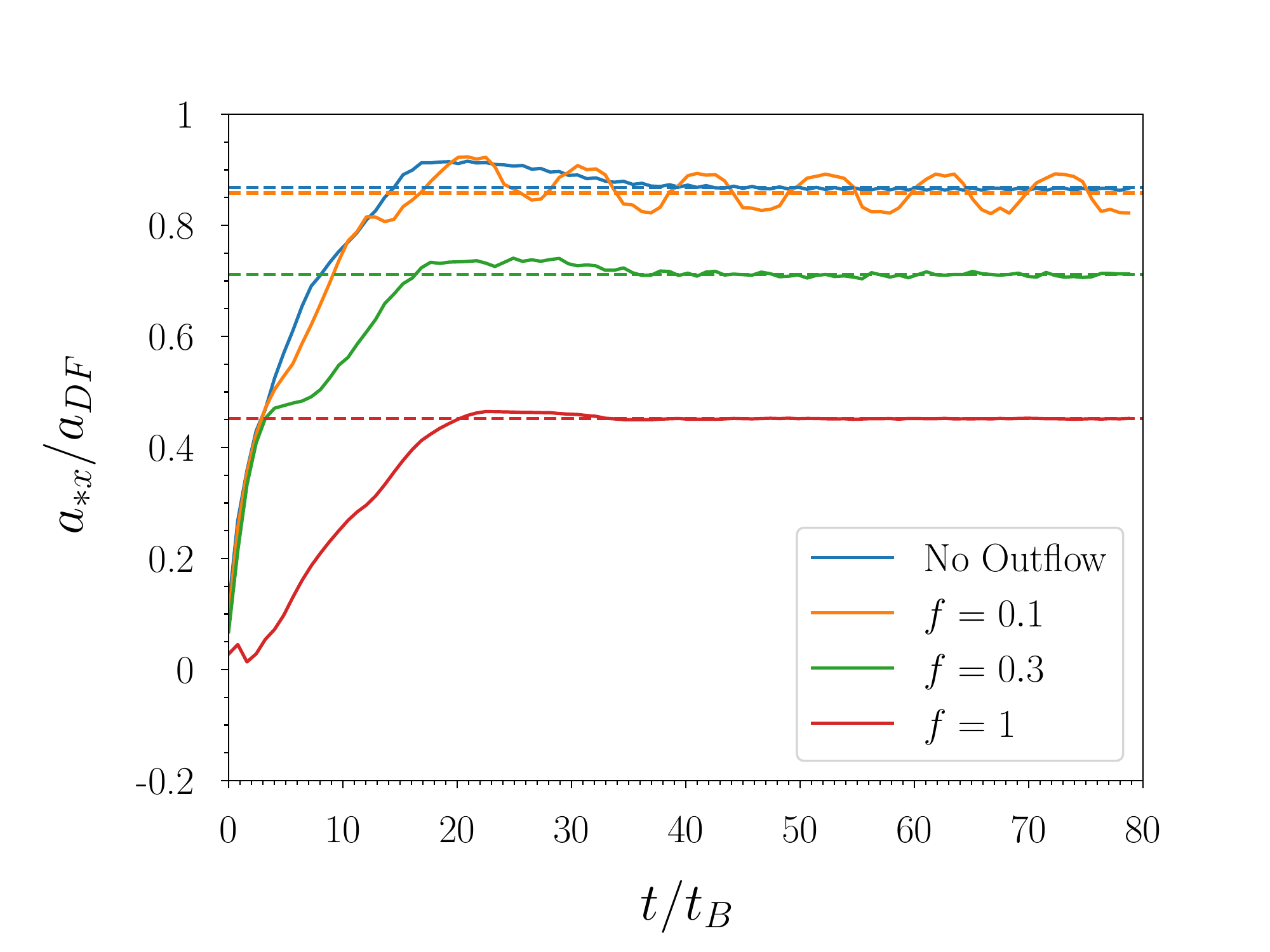}
    \caption{Varying the outflow efficiency $f$ for simulations with outflow in the $z$-direction with $u=0.1$ ($V_w=3\times 10^3$~km/s). \textbf{Upper panel:} the mass accretion rate. \textbf{Lower panel:} acceleration of the star in the $x$ direction. The dashed lines show the average value for $t>30t_B$. The $f=1$ line is not shown in the upper panel since all the accreted gas is ejected leaving a zero accretion rate.}
    \label{fig:varyf_jetz}
\end{figure}

\begin{figure}
    \centering
    \includegraphics[width=.45\textwidth]{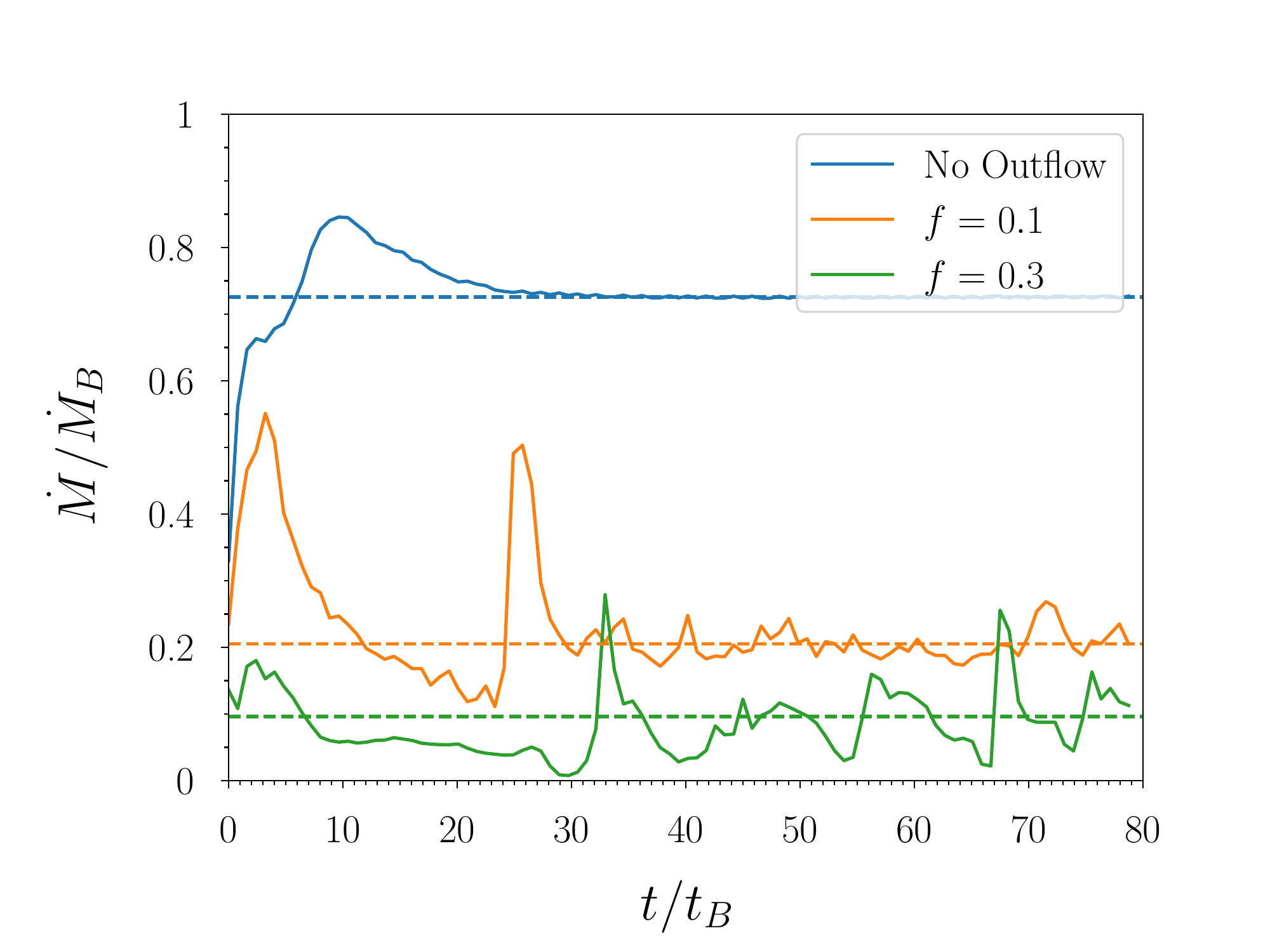}
    \includegraphics[width=.45\textwidth]{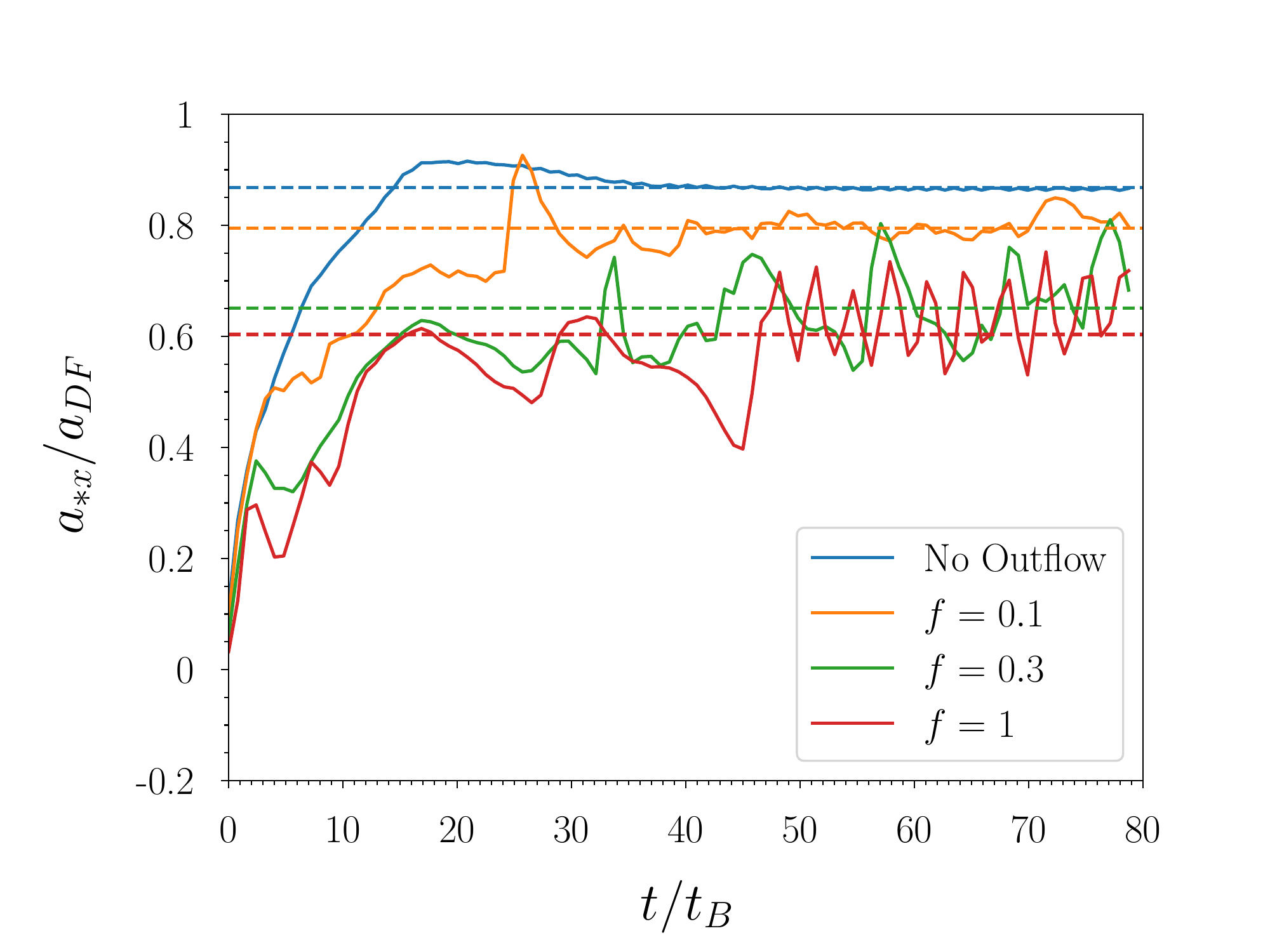}
    \caption{Varying the outflow efficiency $f$ for simulations with outflow in the $x$-direction with $u=0.1$ ($V_w=3\times 10^3$~km/s). \textbf{Upper panel:} the mass accretion rate. \textbf{Lower panel:} acceleration of the star in the $x$ direction. The dashed lines show the average value for $t>30t_B$.The $f=1$ line is not shown in the upper panel since all the accreted gas is ejected leaving a zero accretion rate.}
    \label{fig:varyf_jetx}
\end{figure}

Figures \ref{fig:varyf_jetz} and \ref{fig:varyf_jetx} show how the accretion rate and acceleration can change with the outflow efficiency $f$.
With increasing $f$, both the mass accretion rate and the acceleration decrease whether the jet is along $x$ or $z$ axis.
When the outflow is directed along the $x$ axis, the accretion rate and acceleration are still irregular.
The maximum outflow cases ($f=1$) eject all gas accreted and shut off the accretion; for this reason they are not shown in the plots.
Even with the maximum outflow efficiency, the acceleration decreases but still does not reverse direction.

\subsection{Dynamical Friction} \label{sec:DynFrict}

In the previous section, we have shown that the existence of outflow will reduce total acceleration the compact object experiences.
However, unlike the isotropic case where accretion is totally shut off by the outflow, the ambient gas keeps accreting to the central object.
In our simulation where we set up a background with constant gas density, the overdense wake behind the object extends all the way to the boundary of the computational box.
Due to the long range nature of gravity, the value of total gravitational drag from the overdense wake increases with distance.
The net gravitational acceleration contains two contributions: one from the overdense wake, which diverges with distance; and the other from the underdense region swept by the jet, which points to the opposite direction but converges with increasing box size.

\begin{figure}
    \centering
    \includegraphics[width=.45\textwidth]{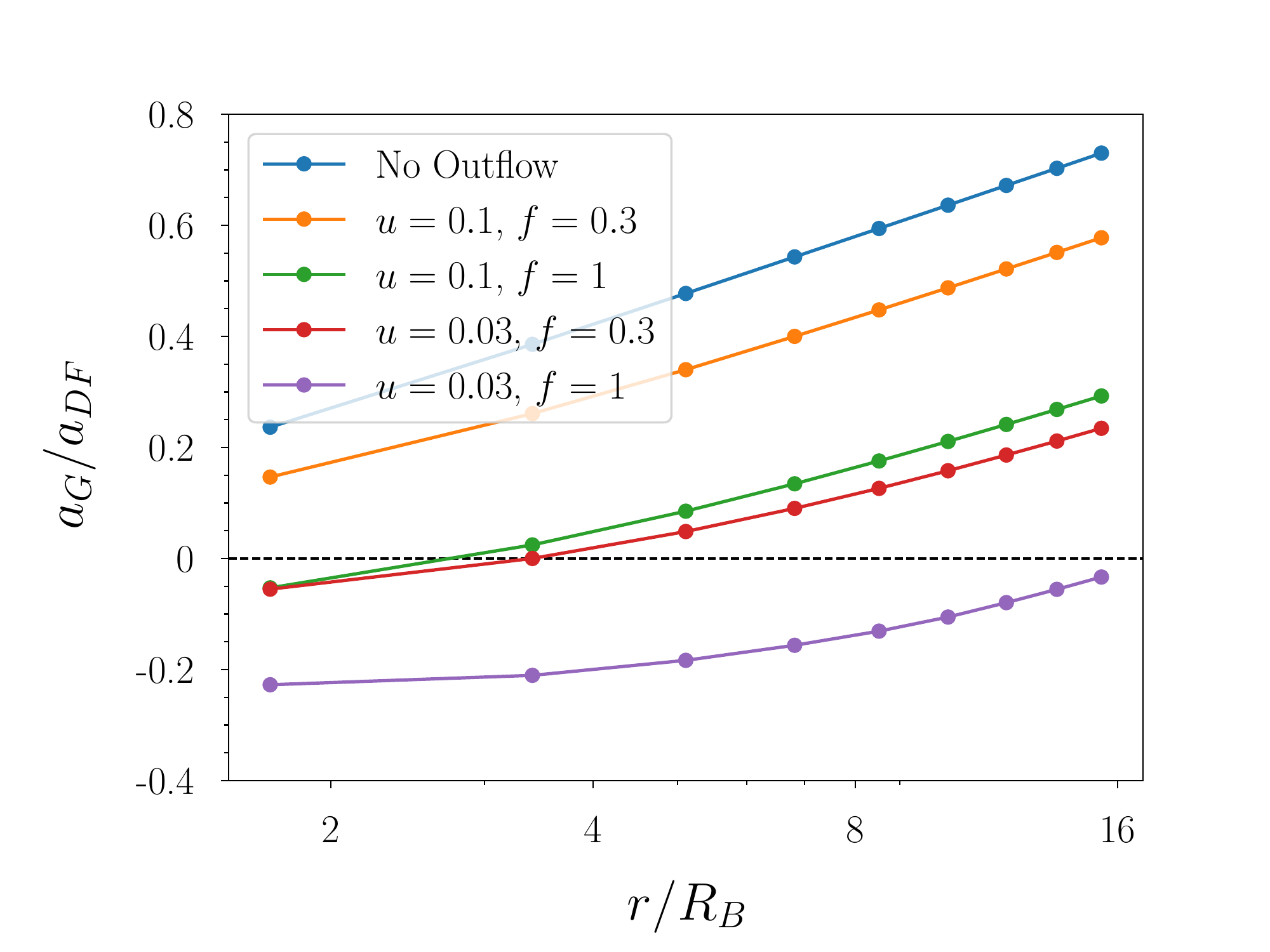}
    \includegraphics[width=.45\textwidth]{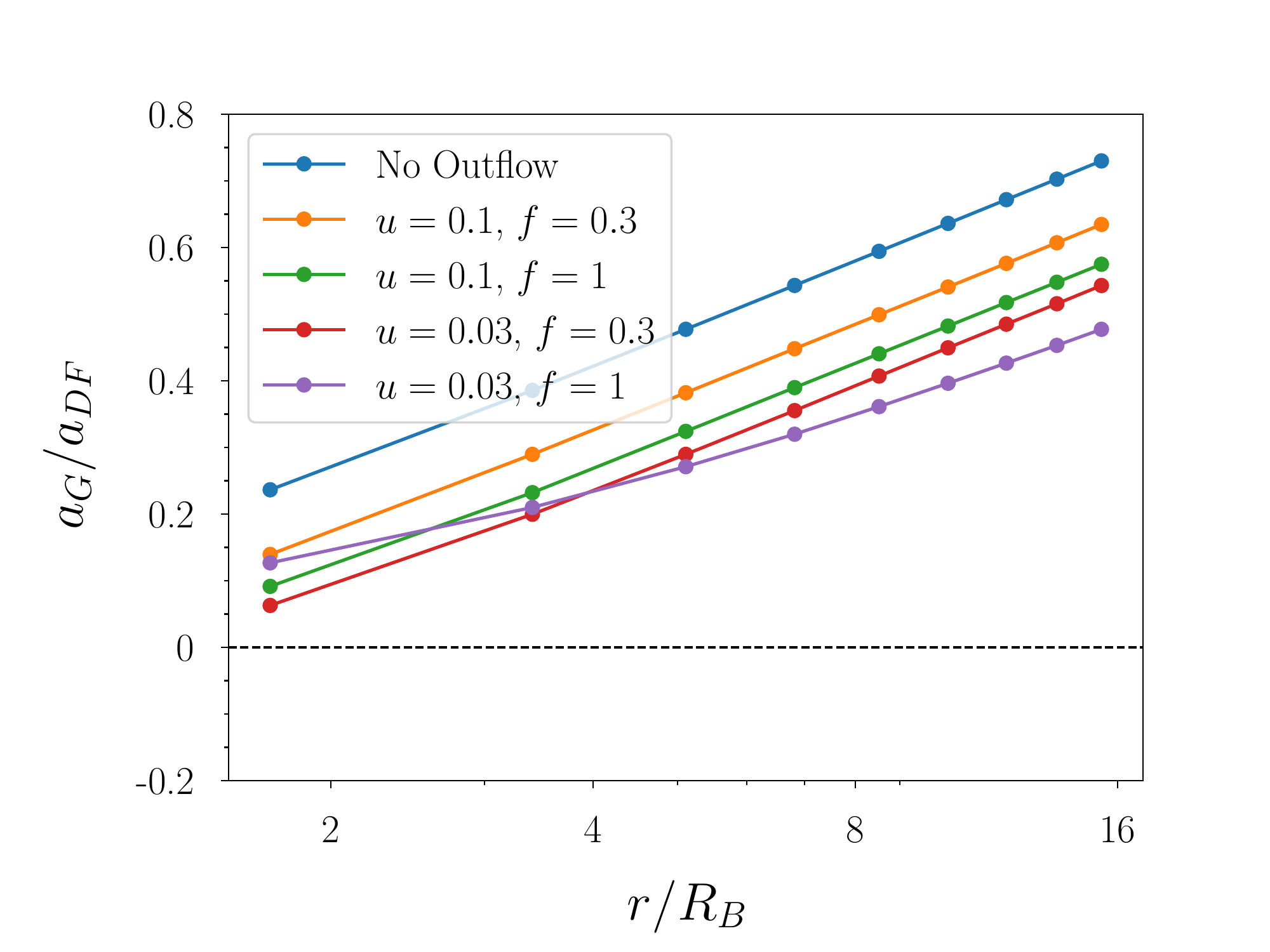}
    \caption{Gravitational acceleration (not including accretion) in the $x$-direction computed from gas within a sphere with radius $r$. The acceleration is the average value for $t>0.01$~yr. \textbf{Upper panel:} the jet is along the $z$ axis, perpendicular to the flow. \textbf{Lower panel:} the jet is along $x$ axis, parallel to the flow.}
    \label{fig:ax_box}
\end{figure}

To demonstrate this point, Figure~\ref{fig:ax_box} shows the amount of gravitational acceleration $a_{G}$ from gas inside a sphere with radius $r$ from the object for different cases.
When the outflow is absent, the analytical dynamical friction is proportional to the logarithm $a_{G}\propto \ln \Lambda$.
As seen in Figure~\ref{fig:ax_box}, the blue line increases logarithmically with the radius $r$, as predicted by the analytical expression.

\begin{figure}
    \centering
    \includegraphics[width=.45\textwidth]{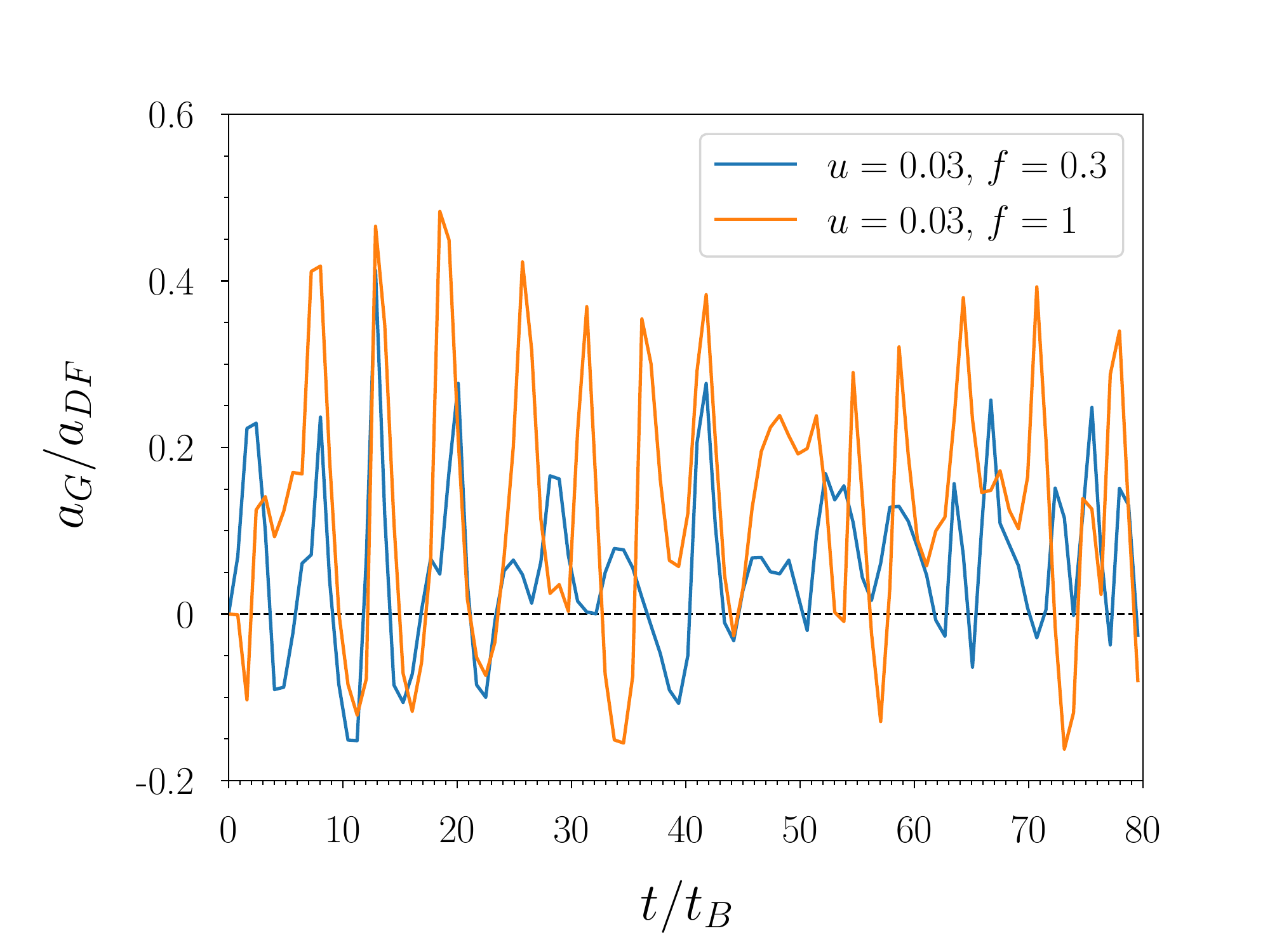}
    \caption{Evolution of gravitational acceleration in the $x$-direction computed from gas within a sphere of radius $r=5\times 10^{11}$~cm for the case where the jet is along $x$ axis.}
    \label{fig:ax_jetx_smallr}
\end{figure}

The same trend of increasing gravitational drag is observed for both cases where the jet is along the $z$ or $x$ axis (perpendicular or parallel to the flow, respectively).
For large $r$, $a_G$ grows logarithmically, signifying the domination of the gravitational drag from the overdense wake.
The effect of outflow is more significant for smaller $r$, especially for powerful jet with small $u$ (large $V_w$) and high efficiency $f$. 
We observe that the gravitational drag from a small sphere of gas $r\leq 4R_B$ points to the same direction as the motion of the object when the jet is along $z$ axis.
For the cases where the jet is along $x$ axis, the gravitational acceleration are all positive.
However, there is significant variability, as shown in Figure~\ref{fig:ax_jetx_smallr}.
The variable acceleration does sometimes become negative of order $0.2a_{DF}$, comparable with the value observed when the jet is along $z$ axis.
When the $f$ is maximal, even though the average value of $a_G$ does not change much, but $a_G$ oscillates more wildly and gives rise to larger negative transient values.

With these results in hand, we reconsider the isotropic wind case  considered in \S~\ref{sec:isotropic}.   
We note that, at sufficiently large distances, the positive force from the gravitational wake should overwhelm the negative dynamical friction observed the simulation presented in \S~\ref{sec:isotropic}.  
This effect was not included in the analysis by \citet{2019arXiv190601186G}, as this was based on  \citet{1996ApJ...459L..31W}'s mass distribution.  

However, the underlying assumption of a uniform background also breaks down on large scales, because self-gravity would cause collapse on a time $\sim (G\rho)^{-1/2}$.  (However this cannot occur in our simulations, thanks to the periodic nature of the self-gravity.) 
For the background fluid to be hydrostatic, therefore, its initial pressure scale height can be at most $\sim c_s (G\rho)^{-1/2}$.  
Significantly, in the astrophysical situations considered in \citet{2019arXiv190601186G}, the spatial scale of the system was not that much greater than the Bondi radius. The finite size is clearly important both for the stellar mass black hole inspiraling through the common envelope, as well as for the stellar or intermediate-mass black hole embedded inside a thin AGN disc. The sign of the dynamical friction in these systems cannot be inferred from the numerical simulations in this paper, and requires a separate investigation.

\section{Discussion}\label{sec:discussion}
We have performed hydrodynamical simulations of a compact object with an outflow moving through ambient gas of uniform density, focusing on the effects of the outflow and accretion on the dynamical friction.
We considered the case of a constant, isotropic wind, the case of no outflow, and the case of a beamed jet with  mass flux proportional to the accretion rate (either perpendicular or parallel to the flow direction).  
Our main results can be summarized as follows: 
\begin{enumerate}
\item Our simulations with strong isotropic winds lead to negative dynamical friction, as predicted by  \citet{2019arXiv190601186G}.
The position of the bow shock and the strength of the negative dynamical friction are in agreement with analytical estimates.
We anticipate that the net force is positive on sufficiently large scales, due to the gravitational wake; however the assumption of uniform density is inconsistent at large scales.
\item Feedback from accretion-powered jets from the compact object is found to significantly reduce the accretion rate and the strength of dynamical friction in the Bondi-Hoyle-Lyttleton picture.
The net force was never negative in our simulations when gravity of all gas from a large simulation domain was accounted for. 
However, the sign was reversed in some simulations, when gravity of the gas only within several Bondi radii was accounted for. 
This implies that the sign of the frictional force on compact objects with outflows needs to be treated with some caution, especially in dense environments. 
\end{enumerate}

We focus on the cases where the outflow is very powerful and the dynamical effect is most visible.
As the outflow power is reduced, either with smaller velocity or smaller $f$, the dynamical effect will also be reduced and asymptote to the case with no outflow as seen in figures presented in Section~\ref{sec:jet}.

Our simulations employed the simplest setup of a uniform background that extends to large distances (formally to infinity). 
Self-gravity of the uniform ambient gas was ignored. In physical systems, finite-size effects, density or velocity gradients, and self-gravity may all be important. Further complexities arise when the accreting object is itself orbiting within a disc of gas (in an AGN or protoplanetary disc), as in that situation forces are exerted at resonances whose position depends upon on the disc rotation profile. 
Our simulations should therefore be interpreted as local models (e.g. in the common envelope) that are approximately valid on scales where density gradients and other effects can be neglected.
Though an outflow generally reduces dynamical friction in our simulations, whether dynamical friction can become negative depends on the competition between the negative effect of the outflow (primarily at small distances) and the over-dense gravitational wake (at the largest distances).
For a specific system, ad hoc simulations are required to determine the exact matter distribution at large distances.

In our simulations with accretion-powered jets, we have explored both cases where the jet is perpendicular and parallel to the object's direction of motion.
The perpendicular case shows a stable accretion and acceleration while the parallel case is more irregular and the accretion is strongly intermittent.
In reality, the direction of the jet can be more irregular as the relative orientation between spin axis of the compact object and its velocity vector in the ambient gas' frame can change during its motion.  
Irregularities in accretion rate and acceleration should be general features when the accretion-powered outflow is present.

In this paper, we assume the compact will accrete matter at the Bondi accretion rate $\dot{M}_B$ if no outflow is active.
A potential worry is that, in all our simulations, $\dot{M}_B$ exceeds the Eddington luminosity, making the accretion hypercritical.
We note, however, that the Eddington limit becomes irrelevant at sufficiently high accretion rate as the photons are trapped and advected inward with the flow \citep{1978PhyS...17..193R,1979MNRAS.187..237B,1986ApJ...308..755B}.
As an example, \citet{2014Sci...343.1330S} report an accreting black hole in M83 with super-Eddington kinetic power.
\citet{1998ApJ...506..780B} suggest that $\dot{M}_B$ is the appropriate accretion rate for compacts in the common envelope phase.
However, population synthesis of binary compact object systems \citep{2002ApJ...572..407B,2007ApJ...662..504B} indicates that accretion at $\dot{M}_B$ would cause neutron stars to collapse to black holes, which would be incompatible with observations.
Hence, the true accretion rate must be reduced \citep{2008ApJ...672L..41R}, and the exact process is still not well understood.
Our simulations show that the dynamical effects of outflow can reduce the true accretion rate below $\dot{M}_B$ which might serve as a partial solution to this problem.

In this work we did not model the outflow launching mechanism in any detail and rather employed a phenomenological model. For a compact object of stellar mass, the outflow launching mechanism acts on scales much smaller than the dynamical scales that we considered, making simultaneous consideration of outflow launching and dynamical friction highly challenging.
How the large and small scale dynamics are coupled and how the large scale dynamics can change the disc formation and jet launching remain interesting questions to be explored.

\section*{Acknowledgements}

We thank Andrei Gruzinov for useful discussions.
PC is supported by the NASA ATP program through NASA grant NNH17ZDA001N-ATP and the Simons Foundation.
We also use the yt software platform for the analysis of the data and generation of plots in this work \citep{yt}.
Research at Perimeter Institute is supported in part by the Government of Canada through the Department of Innovation, Science and Economic Development Canada and by the Province of Ontario through the Ministry of Colleges and Universities.
The Flatiron Institute is supported by the Simons Foundation.



\bibliographystyle{mnras}
\bibliography{ms} 

\begin{thebibliography}{}
\makeatletter
\relax
\def\mn@urlcharsother{\let\do\@makeother \do\$\do\&\do\#\do\^\do\_\do\%\do\~}
\def\mn@doi{\begingroup\mn@urlcharsother \@ifnextchar [ {\mn@doi@}
  {\mn@doi@[]}}
\def\mn@doi@[#1]#2{\def\@tempa{#1}\ifx\@tempa\@empty \href
  {http://dx.doi.org/#2} {doi:#2}\else \href {http://dx.doi.org/#2} {#1}\fi
  \endgroup}
\def\mn@eprint#1#2{\mn@eprint@#1:#2::\@nil}
\def\mn@eprint@arXiv#1{\href {http://arxiv.org/abs/#1} {{\tt arXiv:#1}}}
\def\mn@eprint@dblp#1{\href {http://dblp.uni-trier.de/rec/bibtex/#1.xml}
  {dblp:#1}}
\def\mn@eprint@#1:#2:#3:#4\@nil{\def\@tempa {#1}\def\@tempb {#2}\def\@tempc
  {#3}\ifx \@tempc \@empty \let \@tempc \@tempb \let \@tempb \@tempa \fi \ifx
  \@tempb \@empty \def\@tempb {arXiv}\fi \@ifundefined
  {mn@eprint@\@tempb}{\@tempb:\@tempc}{\expandafter \expandafter \csname
  mn@eprint@\@tempb\endcsname \expandafter{\@tempc}}}

\bibitem[\protect\citeauthoryear{{Begelman}}{{Begelman}}{1979}]{1979MNRAS.187..237B}
{Begelman} M.~C.,  1979, \mn@doi [\mnras] {10.1093/mnras/187.2.237}, \href
  {https://ui.adsabs.harvard.edu/abs/1979MNRAS.187..237B} {187, 237}

\bibitem[\protect\citeauthoryear{{Belczynski}, {Kalogera}  \&
  {Bulik}}{{Belczynski} et~al.}{2002}]{2002ApJ...572..407B}
{Belczynski} K.,  {Kalogera} V.,   {Bulik} T.,  2002, \mn@doi [\apj]
  {10.1086/340304}, \href
  {https://ui.adsabs.harvard.edu/abs/2002ApJ...572..407B} {572, 407}

\bibitem[\protect\citeauthoryear{{Belczynski}, {Taam}, {Kalogera}, {Rasio}  \&
  {Bulik}}{{Belczynski} et~al.}{2007}]{2007ApJ...662..504B}
{Belczynski} K.,  {Taam} R.~E.,  {Kalogera} V.,  {Rasio} F.~A.,   {Bulik} T.,
  2007, \mn@doi [\apj] {10.1086/513562}, \href
  {https://ui.adsabs.harvard.edu/abs/2007ApJ...662..504B} {662, 504}

\bibitem[\protect\citeauthoryear{{Bethe} \& {Brown}}{{Bethe} \&
  {Brown}}{1998}]{1998ApJ...506..780B}
{Bethe} H.~A.,  {Brown} G.~E.,  1998, \mn@doi [\apj] {10.1086/306265}, \href
  {https://ui.adsabs.harvard.edu/abs/1998ApJ...506..780B} {506, 780}

\bibitem[\protect\citeauthoryear{{Bleuler} \& {Teyssier}}{{Bleuler} \&
  {Teyssier}}{2014}]{2014MNRAS.445.4015B}
{Bleuler} A.,  {Teyssier} R.,  2014, \mn@doi [\mnras] {10.1093/mnras/stu2005},
  \href {http://adsabs.harvard.edu/abs/2014MNRAS.445.4015B} {445, 4015}

\bibitem[\protect\citeauthoryear{{Blondin}}{{Blondin}}{1986}]{1986ApJ...308..755B}
{Blondin} J.~M.,  1986, \mn@doi [\apj] {10.1086/164548}, \href
  {https://ui.adsabs.harvard.edu/abs/1986ApJ...308..755B} {308, 755}

\bibitem[\protect\citeauthoryear{{Blondin} \& {Pope}}{{Blondin} \&
  {Pope}}{2009}]{2009ApJ...700...95B}
{Blondin} J.~M.,  {Pope} T.~C.,  2009, \mn@doi [\apj]
  {10.1088/0004-637X/700/1/95}, \href
  {https://ui.adsabs.harvard.edu/abs/2009ApJ...700...95B} {700, 95}

\bibitem[\protect\citeauthoryear{{Blondin} \& {Raymer}}{{Blondin} \&
  {Raymer}}{2012}]{2012ApJ...752...30B}
{Blondin} J.~M.,  {Raymer} E.,  2012, \mn@doi [\apj]
  {10.1088/0004-637X/752/1/30}, \href
  {https://ui.adsabs.harvard.edu/abs/2012ApJ...752...30B} {752, 30}

\bibitem[\protect\citeauthoryear{{Blondin}, {Kallman}, {Fryxell}  \&
  {Taam}}{{Blondin} et~al.}{1990}]{1990ApJ...356..591B}
{Blondin} J.~M.,  {Kallman} T.~R.,  {Fryxell} B.~A.,   {Taam} R.~E.,  1990,
  \mn@doi [\apj] {10.1086/168865}, \href
  {https://ui.adsabs.harvard.edu/abs/1990ApJ...356..591B} {356, 591}

\bibitem[\protect\citeauthoryear{{Bondi}}{{Bondi}}{1952}]{1952MNRAS.112..195B}
{Bondi} H.,  1952, \mn@doi [\mnras] {10.1093/mnras/112.2.195}, \href
  {https://ui.adsabs.harvard.edu/abs/1952MNRAS.112..195B} {112, 195}

\bibitem[\protect\citeauthoryear{{Bondi} \& {Hoyle}}{{Bondi} \&
  {Hoyle}}{1944}]{1944MNRAS.104..273B}
{Bondi} H.,  {Hoyle} F.,  1944, \mn@doi [\mnras] {10.1093/mnras/104.5.273},
  \href {https://ui.adsabs.harvard.edu/abs/1944MNRAS.104..273B} {104, 273}

\bibitem[\protect\citeauthoryear{{Chandrasekhar}}{{Chandrasekhar}}{1943}]{1943ApJ....97..255C}
{Chandrasekhar} S.,  1943, \mn@doi [\apj] {10.1086/144517}, \href
  {https://ui.adsabs.harvard.edu/abs/1943ApJ....97..255C} {97, 255}

\bibitem[\protect\citeauthoryear{{Dubois}, {Devriendt}, {Slyz}  \&
  {Teyssier}}{{Dubois} et~al.}{2010}]{2010MNRAS.409..985D}
{Dubois} Y.,  {Devriendt} J.,  {Slyz} A.,   {Teyssier} R.,  2010, \mn@doi
  [\mnras] {10.1111/j.1365-2966.2010.17338.x}, \href
  {http://adsabs.harvard.edu/abs/2010MNRAS.409..985D} {409, 985}

\bibitem[\protect\citeauthoryear{{Edgar}}{{Edgar}}{2004}]{2004NewAR..48..843E}
{Edgar} R.,  2004, \mn@doi [\nar] {10.1016/j.newar.2004.06.001}, \href
  {https://ui.adsabs.harvard.edu/abs/2004NewAR..48..843E} {48, 843}

\bibitem[\protect\citeauthoryear{{Fabrika}}{{Fabrika}}{2004}]{2004ASPRv..12....1F}
{Fabrika} S.,  2004, \apspr, \href
  {https://ui.adsabs.harvard.edu/abs/2004ASPRv..12....1F} {12, 1}

\bibitem[\protect\citeauthoryear{{Fryxell} \& {Taam}}{{Fryxell} \&
  {Taam}}{1988}]{1988ApJ...335..862F}
{Fryxell} B.~A.,  {Taam} R.~E.,  1988, \mn@doi [\apj] {10.1086/166973}, \href
  {https://ui.adsabs.harvard.edu/abs/1988ApJ...335..862F} {335, 862}

\bibitem[\protect\citeauthoryear{{Gruzinov}, {Levin}  \& {Matzner}}{{Gruzinov}
  et~al.}{2019}]{2019arXiv190601186G}
{Gruzinov} A.,  {Levin} Y.,   {Matzner} C.~D.,  2019, arXiv e-prints, \href
  {https://ui.adsabs.harvard.edu/abs/2019arXiv190601186G} {p. arXiv:1906.01186}

\bibitem[\protect\citeauthoryear{{Hillel}, {Schreier}  \& {Soker}}{{Hillel}
  et~al.}{2019}]{2019arXiv191204662H}
{Hillel} S.,  {Schreier} R.,   {Soker} N.,  2019, arXiv e-prints, \href
  {https://ui.adsabs.harvard.edu/abs/2019arXiv191204662H} {p. arXiv:1912.04662}

\bibitem[\protect\citeauthoryear{{Hoyle} \& {Lyttleton}}{{Hoyle} \&
  {Lyttleton}}{1939}]{1939PCPS...35..405H}
{Hoyle} F.,  {Lyttleton} R.~A.,  1939, \mn@doi [Proceedings of the Cambridge
  Philosophical Society] {10.1017/S0305004100021150}, \href
  {https://ui.adsabs.harvard.edu/abs/1939PCPS...35..405H} {35, 405}

\bibitem[\protect\citeauthoryear{{Kley}, {Shankar}  \& {Burkert}}{{Kley}
  et~al.}{1995}]{1995A&A...297..739K}
{Kley} W.,  {Shankar} A.,   {Burkert} A.,  1995, \aap, \href
  {https://ui.adsabs.harvard.edu/abs/1995A&A...297..739K} {297, 739}

\bibitem[\protect\citeauthoryear{{MacLeod}, {Antoni}, {Murguia-Berthier},
  {Macias}  \& {Ramirez-Ruiz}}{{MacLeod} et~al.}{2017}]{2017ApJ...838...56M}
{MacLeod} M.,  {Antoni} A.,  {Murguia-Berthier} A.,  {Macias} P.,
  {Ramirez-Ruiz} E.,  2017, \mn@doi [\apj] {10.3847/1538-4357/aa6117}, \href
  {https://ui.adsabs.harvard.edu/abs/2017ApJ...838...56M} {838, 56}

\bibitem[\protect\citeauthoryear{{Masset}}{{Masset}}{2017}]{2017MNRAS.472.4204M}
{Masset} F.~S.,  2017, \mn@doi [\mnras] {10.1093/mnras/stx2271}, \href
  {https://ui.adsabs.harvard.edu/abs/2017MNRAS.472.4204M} {472, 4204}

\bibitem[\protect\citeauthoryear{{Masset} \& {Velasco Romero}}{{Masset} \&
  {Velasco Romero}}{2017}]{2017MNRAS.465.3175M}
{Masset} F.~S.,  {Velasco Romero} D.~A.,  2017, \mn@doi [\mnras]
  {10.1093/mnras/stw3008}, \href
  {https://ui.adsabs.harvard.edu/abs/2017MNRAS.465.3175M} {465, 3175}

\bibitem[\protect\citeauthoryear{{Murray}, {Chang}, {Murray}  \&
  {Pittman}}{{Murray} et~al.}{2017}]{2017MNRAS.465.1316M}
{Murray} D.~W.,  {Chang} P.,  {Murray} N.~W.,   {Pittman} J.,  2017, \mn@doi
  [\mnras] {10.1093/mnras/stw2796}, \href
  {http://adsabs.harvard.edu/abs/2017MNRAS.465.1316M} {465, 1316}

\bibitem[\protect\citeauthoryear{{Murray}, {Goyal}  \& {Chang}}{{Murray}
  et~al.}{2018}]{2018MNRAS.475.1023M}
{Murray} D.,  {Goyal} S.,   {Chang} P.,  2018, \mn@doi [\mnras]
  {10.1093/mnras/stx3153}, \href
  {https://ui.adsabs.harvard.edu/abs/2018MNRAS.475.1023M} {475, 1023}

\bibitem[\protect\citeauthoryear{{Ostriker}}{{Ostriker}}{1999}]{1999ApJ...513..252O}
{Ostriker} E.~C.,  1999, \mn@doi [\apj] {10.1086/306858}, \href
  {https://ui.adsabs.harvard.edu/abs/1999ApJ...513..252O} {513, 252}

\bibitem[\protect\citeauthoryear{{Park} \& {Bogdanovi{\'c}}}{{Park} \&
  {Bogdanovi{\'c}}}{2017}]{2017ApJ...838..103P}
{Park} K.,  {Bogdanovi{\'c}} T.,  2017, \mn@doi [\apj]
  {10.3847/1538-4357/aa65ce}, \href
  {https://ui.adsabs.harvard.edu/abs/2017ApJ...838..103P} {838, 103}

\bibitem[\protect\citeauthoryear{{Park} \& {Ricotti}}{{Park} \&
  {Ricotti}}{2013}]{2013ApJ...767..163P}
{Park} K.,  {Ricotti} M.,  2013, \mn@doi [\apj] {10.1088/0004-637X/767/2/163},
  \href {https://ui.adsabs.harvard.edu/abs/2013ApJ...767..163P} {767, 163}

\bibitem[\protect\citeauthoryear{{Rees}}{{Rees}}{1978}]{1978PhyS...17..193R}
{Rees} M.~J.,  1978, \mn@doi [\physscr] {10.1088/0031-8949/17/3/010}, \href
  {https://ui.adsabs.harvard.edu/abs/1978PhyS...17..193R} {17, 193}

\bibitem[\protect\citeauthoryear{{Ricker} \& {Taam}}{{Ricker} \&
  {Taam}}{2008}]{2008ApJ...672L..41R}
{Ricker} P.~M.,  {Taam} R.~E.,  2008, \mn@doi [\apjl] {10.1086/526343}, \href
  {https://ui.adsabs.harvard.edu/abs/2008ApJ...672L..41R} {672, L41}

\bibitem[\protect\citeauthoryear{{Rosdahl}, {Blaizot}, {Aubert}, {Stranex}  \&
  {Teyssier}}{{Rosdahl} et~al.}{2013}]{2013MNRAS.436.2188R}
{Rosdahl} J.,  {Blaizot} J.,  {Aubert} D.,  {Stranex} T.,   {Teyssier} R.,
  2013, \mn@doi [\mnras] {10.1093/mnras/stt1722}, \href
  {https://ui.adsabs.harvard.edu/abs/2013MNRAS.436.2188R} {436, 2188}

\bibitem[\protect\citeauthoryear{{Ruffert}}{{Ruffert}}{1997}]{1997A&A...317..793R}
{Ruffert} M.,  1997, \aap, \href
  {https://ui.adsabs.harvard.edu/abs/1997A&A...317..793R} {317, 793}

\bibitem[\protect\citeauthoryear{{Ruffert}}{{Ruffert}}{1999}]{1999A&A...346..861R}
{Ruffert} M.,  1999, \aap, \href
  {https://ui.adsabs.harvard.edu/abs/1999A&A...346..861R} {346, 861}

\bibitem[\protect\citeauthoryear{{Shiber}, {Iaconi}, {De Marco}  \&
  {Soker}}{{Shiber} et~al.}{2019}]{2019MNRAS.488.5615S}
{Shiber} S.,  {Iaconi} R.,  {De Marco} O.,   {Soker} N.,  2019, \mn@doi
  [\mnras] {10.1093/mnras/stz2013}, \href
  {https://ui.adsabs.harvard.edu/abs/2019MNRAS.488.5615S} {488, 5615}

\bibitem[\protect\citeauthoryear{{Shima}, {Matsuda}, {Takeda}  \&
  {Sawada}}{{Shima} et~al.}{1985}]{1985MNRAS.217..367S}
{Shima} E.,  {Matsuda} T.,  {Takeda} H.,   {Sawada} K.,  1985, \mn@doi [\mnras]
  {10.1093/mnras/217.2.367}, \href
  {https://ui.adsabs.harvard.edu/abs/1985MNRAS.217..367S} {217, 367}

\bibitem[\protect\citeauthoryear{{Soker}}{{Soker}}{2016}]{2016NewAR..75....1S}
{Soker} N.,  2016, \mn@doi [\nar] {10.1016/j.newar.2016.08.002}, \href
  {https://ui.adsabs.harvard.edu/abs/2016NewAR..75....1S} {75, 1}

\bibitem[\protect\citeauthoryear{{Soria}, {Long}, {Blair}, {Godfrey}, {Kuntz},
  {Lenc}, {Stockdale}  \& {Winkler}}{{Soria}
  et~al.}{2014}]{2014Sci...343.1330S}
{Soria} R.,  {Long} K.~S.,  {Blair} W.~P.,  {Godfrey} L.,  {Kuntz} K.~D.,
  {Lenc} E.,  {Stockdale} C.,   {Winkler} P.~F.,  2014, \mn@doi [Science]
  {10.1126/science.1248759}, \href
  {https://ui.adsabs.harvard.edu/abs/2014Sci...343.1330S} {343, 1330}

\bibitem[\protect\citeauthoryear{{Souza Lima}, {Mayer}, {Capelo}  \&
  {Bellovary}}{{Souza Lima} et~al.}{2017}]{2017ApJ...838...13S}
{Souza Lima} R.,  {Mayer} L.,  {Capelo} P.~R.,   {Bellovary} J.~M.,  2017,
  \mn@doi [\apj] {10.3847/1538-4357/aa5d19}, \href
  {https://ui.adsabs.harvard.edu/abs/2017ApJ...838...13S} {838, 13}

\bibitem[\protect\citeauthoryear{{Taam}, {Fu}  \& {Fryxell}}{{Taam}
  et~al.}{1991}]{1991ApJ...371..696T}
{Taam} R.~E.,  {Fu} A.,   {Fryxell} B.~A.,  1991, \mn@doi [\apj]
  {10.1086/169935}, \href
  {https://ui.adsabs.harvard.edu/abs/1991ApJ...371..696T} {371, 696}

\bibitem[\protect\citeauthoryear{{Teyssier}}{{Teyssier}}{2002}]{2002A&A...385..337T}
{Teyssier} R.,  2002, \mn@doi [\aap] {10.1051/0004-6361:20011817}, \href
  {https://ui.adsabs.harvard.edu/#abs/2002A&A...385..337T} {385, 337}

\bibitem[\protect\citeauthoryear{{Turk}, {Smith}, {Oishi}, {Skory}, {Skillman},
  {Abel}  \& {Norman}}{{Turk} et~al.}{2011}]{yt}
{Turk} M.~J.,  {Smith} B.~D.,  {Oishi} J.~S.,  {Skory} S.,  {Skillman} S.~W.,
  {Abel} T.,   {Norman} M.~L.,  2011, \mn@doi [The Astrophysical Journal
  Supplement Series] {10.1088/0067-0049/192/1/9}, \href
  {http://adsabs.harvard.edu/abs/2011ApJS..192....9T} {192, 9}

\bibitem[\protect\citeauthoryear{{Wilkin}}{{Wilkin}}{1996}]{1996ApJ...459L..31W}
{Wilkin} F.~P.,  1996, \mn@doi [\apj] {10.1086/309939}, \href
  {https://ui.adsabs.harvard.edu/abs/1996ApJ...459L..31W} {459, L31}

\makeatother
\end{thebibliography}



\bsp	
\label{lastpage}
\end{document}